\documentclass[12pt]{article}
\usepackage{a4,epsfig}
\begin{document}

%===================> ADD here your LATEX definitions

%================================================ ========================% 

%################################################## titlepage declaration

\begin{titlepage}

\pagenumbering{arabic}
\vspace*{-1.5cm}
\begin{tabular*}{15.cm}{lc@{\extracolsep{\fill}}r}
%{\bf DELPHI Collaboration} & 
%\hspace*{1.3cm} \epsfig{figure=/afs/cern.ch/user/d/delnote/tex/dolphin_bw.eps,width=1.2cm,height=1.2cm}
&
%===================> DELPHI note number       =====> To be filled <=====%
%DELPHI 2016-001 PHYS XXX   
%========================================================================%
\\
& &
%===================> DELPHI note date         =====> To be filled <=====%
10 January 2020
%========================================================================%
\\
&&\\ \hline
\end{tabular*}
\vspace*{1.0cm}
\begin{center}
\Large 
{\bf \boldmath
%===================> DELPHI note title        =====> To be filled <=====%
A search for anomalous Cherenkov rings
%========================================================================%
}

\vspace*{1.0cm}
\normalsize { 
%===================> DELPHI note author list  =====> To be filled <=====%
   {\bf V. F. Perepelitsa}\\
   {\footnotesize ITEP, Moscow            }\\ 
   
   {\bf T. Ekelof}\\
   {\footnotesize Department of Physics and Astronomy, Uppsala University}\\
   
   {\bf A. Ferrer}\\
   {\footnotesize IFIC, Valencia University}\\

   {\bf B. R. French}\\
   {\footnotesize bernardfrench@bluewin.ch}\\
%========================================================================%
}
%\vspace*{0.4cm}
\end{center}
\vspace{\fill}
\begin{abstract}
\noindent
%===================> DELPHI note abstract     =====> To be filled <=====%
The results of a search with the DELPHI Barrel RICH for Cherenkov rings 
having radii greater than those produced by ultrarelativistic particles are 
presented. The search for such anomalous rings, associated with tracks
from electron-like particles, is based on the data collected 
by the DELPHI Collaboration at CERN during the LEP1 and LEP2 periods.
The DELPHI RICH detector was conceived for the identification of the stable
and quasi-stable hadrons ($\pi/K/p$). The present analysis, made with the goal
of finding anomalous Cherenkov rings, was aimed at the investigation of
electron-like particles. 
A subsample of events containing anomalous rings
has been identified for which the probability that the reconstructed rings
in a given event are due to fortuitous combinations of background hits is low
($10^{-3}$ or less). A detailed study of background sources
%and possible systematic effects that are 
capable of producing apparently anomalous rings has been done;
it indicates that the background hypothesis has a low probability.
%(less than $2.2\times 10^{-8}$). 
Additional arguments against this hypothesis are provided by a comparison
of rates of events with single and double anomalous rings in the gaseous 
radiator, indicating a clear tendency for associated production of the 
anomalous rings, and by the observation of a high degree of correlation between
anomalous ring radii in the liquid and gaseous radiators in these events.

This work has been performed by the authors following the rules for external
access to the DELPHI archived data, as established in
http://delphiwww.cern.ch/delsec/finalrules/FINALrules011203.pdf
The opinions, findings and conclusions expressed in this material are those
of the authors alone and do not reflect in any way the views of the 
DELPHI Collaboration.
%=========================================================================%
\end{abstract}
%\vspace{\fill}
\end{titlepage}

%\pagebreak

%\begin{titlepage}
%\mbox{}
%\end{titlepage}

%\pagebreak

%\setcounter{page}{1}    

%##################################################################### Text

%==================> DELPHI note text          =====> To be filled <======%
\section{Introduction}
The results of a search for anomalous Cherenkov rings 
produced in the DELPHI detector RICH in the events of $e^+ e^-$ interactions 
recorded in the DELPHI experiment at LEP (CERN) are presented. 
The term ``anomalous rings" here stands for rings 
having radii greater than those produced by ultrarelativistic particles 
which are termed here ``standard rings". In this article standard rings are
defined as having the ring radius compatible with 667~mrad in the liquid 
radiator and 62~mrad in the gaseous radiator, while the anomalous rings are
defined as having the ring radius exceeding 700~mrad in the liquid radiator and
72~mrad in the gaseous radiator.
% \footnote{Thus the minimal$ radii of the 
%anomalous rings are by about 4 s.d. larger than the radii of the standard 
%rings, see Sect.~2.1.}.    

The DELPHI detector RICH was conceived for the identification of the stable
and quasi-stable hadrons ($\pi/K/p$). The present analysis, made with the goal
of finding anomalous Cherenkov rings, is aimed at the investigation of
electron-like particles, which implies enhanced background from electromagnetic
showers. Therefore a careful study of this background has been carried out 
as described below.

The data used consist of about $10^7$ non-hadronic events 
corresponding to an integrated luminosity of 0.76~fb$^{-1}$.

This note is organized as follows. In Section~2 we describe the experimental 
method and present several examples of anomalous ring candidates. 
The topologies of events investigated are described in Section~3. 
Section~4 itemises the selection cuts used in a preliminary selection
of events of the above topologies that are subsequently investigated for a
presence of anomalous rings; it terminates with a description of the criteria
of the final selection of these rings. Section~5 lists the known systematic 
effects which could produce hit patterns that imitate anomalous rings. 
Section~6 describes the studies of physical backgrounds due to statistical 
effects induced by the fortuitous combinations of the hits presented in the 
RICH hit patterns. Sections~7 and 8 describe additional tests of the background
hypothesis. In Sections~9 and 10 further principal arguments against this 
hypothesis are given. 
Sections~11 and 12 provide a summary and conclusion.

There are two Appendices to the paper which are complementary to the main text.
They describe the algorithm used to find Cherenkov rings and the method of 
calculation of the probability that ring, observed in a given 
RICH hit pattern, can be a fortuitous combination of RICH background hits. 

\section{Experimental details}
\subsection{Experimental technique}
The DELPHI detector is described in detail in \cite{delphi1,delphi2}.
The following is a list of the subdetector units relevant for
this analysis: the vertex and inner detectors (VD and ID), the main tracker
%of the DELPHI detector called
(the Time Projection Chamber, TPC), the outer
detector (OD), the barrel electromagnetic calorimeter (the High-density
Projection Chamber, HPC), the hadronic calorimeter (HCAL), and the barrel muon
chambers (MUB). The principal detector used in this analysis was the barrel
Ring Imaging Cherenkov detector (Barrel RICH).

The DELPHI Barrel RICH is described in detail in 
\cite{rich1,rich2,rich3,rich4,rich5} and has demonstrated to have high
performance  for identification of hadrons ($\pi, K, p$).  
The detector contained two radiators, the liquid radiator 
($C_6 F_{14}$, refraction index $n=1.273$) and the gaseous radiator
($C_5 F_{12},~n = 1.00194$). The Cherenkov photons from these two radiators 
were detected by TPC-like photon detectors consisting of 24 pairs of quartz 
drift tubes (drift boxes), covering a full azimuthal range with 24 Barrel 
RICH sectors, extended to $\pm 155$~cm from the mid-wall. 
The drift gas (75\% methane $CH_4$ and 25\% ethane $C_2 H_6$) was doped
with 0.1\% of the photosensitive agent TMAE, by which the ultraviolet
Cherenkov photons were converted into single free photoelectrons, the mean
photon conversion length being equal to 1.8~cm. The TMAE photo-ionization
threshold of 5.63~eV and the transmission cut-off of the liquid radiator and
drift tube quartz windows of 7.50~eV limited the photon wavelength detection
range to the band of 165 - 220~nm.

The principles of the creation of Cherenkov ring images in both radiators of 
the DELPHI Barrel RICH are illustrated in the following figure:
\includegraphics[height=8.0cm,width=18.0cm]{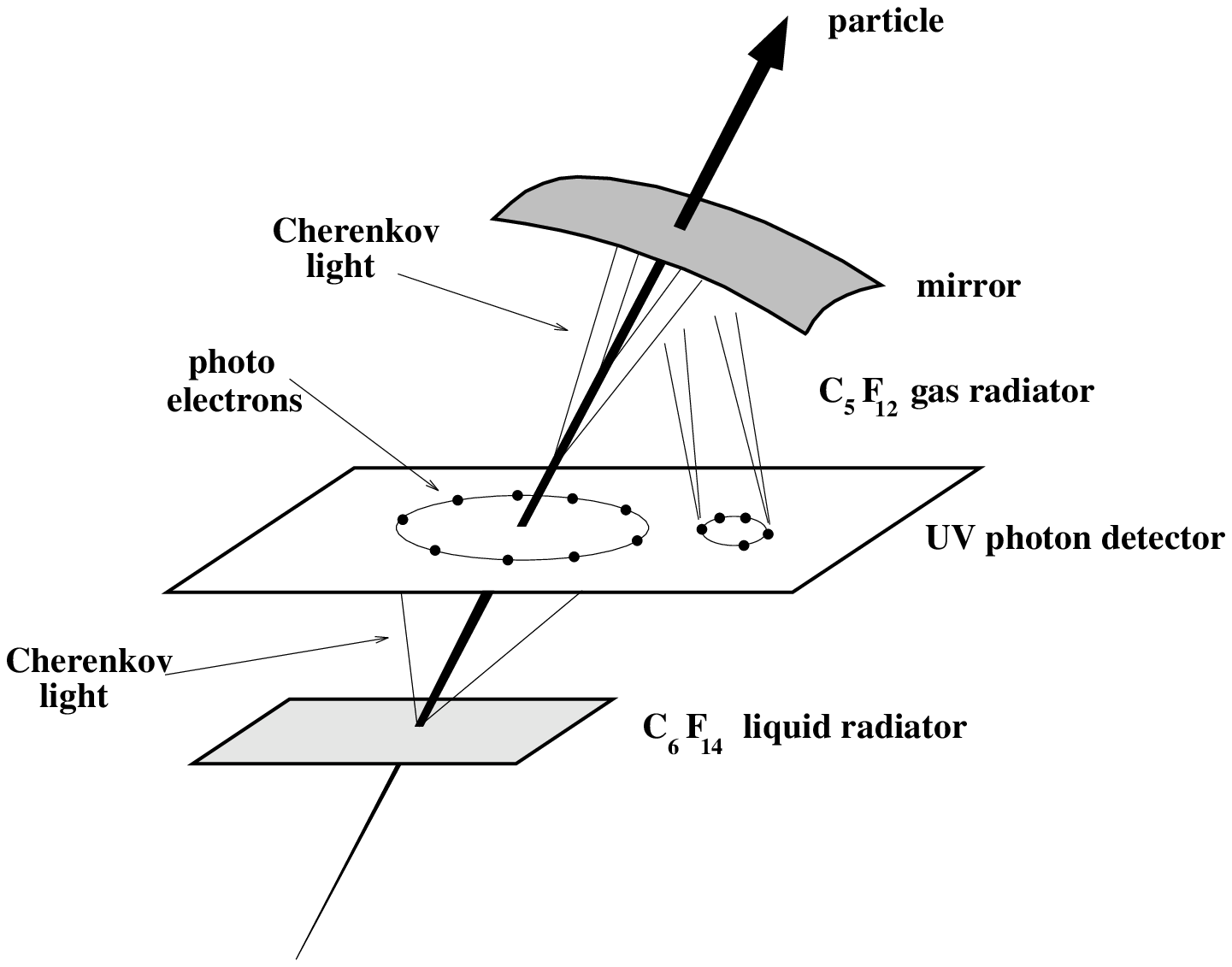}
The rings formed by the hits of photons originating from a given radiator were 
distinguished by the location of the hits in two different parts of the drift 
boxes, separated by the box median plane, the inner part for the liquid and 
outer part for the gaseous radiators (in the present paper ``inner" and 
``outer" positions in the DELPHI detector is with respect to the 
colliding beam axis).

The standard ring radii in the liquid and gas radiators, expressed in angular 
units, were 667 and 62~mrad, respectively. The total single photon standard
error of the liquid radiator $\sigma_{p.e}$ was 18-28~mrad
for standard rings. Each sector of the gaseous radiator was equipped
with six parabolic mirrors distributed along the $z$ axis~\footnote{In the 
DELPHI reference frame the $z$ axis is along the direction of the $e^-$ beam.
It defined the particle polar angles $\Theta$, while the particle azimuthal
angles $\Phi$ were defined in the $xy$ plane.},
which focused the Cherenkov photons generated in the radiator volume back onto
the photon detector. The gas radiator single photon standard error  
$\sigma_{p.e.}$ was about 4~mrad for small radius anomalous rings, increasing 
by an order of magnitude for very big rings due to geometric aberrations 
intrinsic to parabolic mirror optics. With the average Cherenkov photon numbers
per standard ring of 14 and 8~\cite{delphi2}~\footnote{These numbers reduce to 
9 and 5, respectively, within $\pm 1 \sigma_{p.e}$ cited above.} 
for the liquid and gaseous radiators, respectively, the Cherenkov cone angular 
error $\sigma$ for the rings associated with high momentum hadronic tracks
was about 9~mrad for the liquid radiator and about 2.7~mrad for the gaseous 
radiator \cite{kluit}.

After the processing of the raw data with the use of the DELPHI general pattern
recognition package $DELANA$ \cite{delphi2} to produce the DELPHI data summary
tapes (DST's), the RICH data for each individual track consisted of
%two sets of
photon trajectories starting on track segments inside a given radiator
volume, and terminating inside the RICH drift boxes on hits detected in the
latter. This information was treated rather differently
in the standard DELPHI analysis and in the analysis presented here.

The DELPHI standard way was to use the reconstructed Cherenkov angles 
to identify the particle producing the track by applying the maximum 
likelihood technique \cite{delphi2}. Five mass hypotheses 
($m_e,~m_\mu,~m_\pi,~m_K,~m_p$) 
were tried and the results were used to identify the particle. 
Additionally, an independent estimation of the ring radii relevant to each 
charged particle traversing the RICH was made and the results were stored on
the DST's~\cite{kluit}, including the averaged Cherenkov angle, the number of 
photoelectrons in the ring, an estimation of the background, etc., 
%together with information on individual hits, 
and these data were used in this analysis as described below.  Unfortunately, 
for the present investigation, the hits which could correspond to
Cherenkov angles exceeding certain limits were not tried in this procedure,
these limits being 750~mrad for the liquid radiator and 102~mrad for the 
gaseous one.
The information about these large angle hits is not available 
on any of the types of DELPHI DST's. This obviously restricted the power of the
preliminary selection of events containing anomalous rings at the DST level.
Such a selection, described in Sect.~4, was based on other event signatures,
though the information about Cherenkov angles within the limits mentioned above
was used in part for the selection, as will be further explained.

The events tagged by this preliminary selection were reprocessed, and the RICH 
data in them were treated following a procedure specially developed for the 
present analysis. It starts with an analysis of the information stored in raw 
data tapes extracted by the DELPHI event server. Then several steps of the 
standard procedure, described above, are performed, to reconstruct the 
Cherenkov photon directions with respect to tracks,
however without any restrictions on the radiation angles. After this
the photon directions were used, as described below, to find positions of  
the photon impact points on the plane
perpendicular to the track trajectories in the corresponding radiator which 
will here be called the Cherenkov plane. The coordinate frame in this plane has
its origin at the position of the center of the expected ring as reconstructed 
from the track parameters. The plane coordinates $x, y$ of the photon impact 
point, expressed in angular units, are calculated from
the polar and azimuthal angles of the Cherenkov photons, $\theta_c$
and $\phi_c$: $x = \theta_c \sin \phi_c$, $y = \theta_c \cos \phi_c$. 
The resulting hit pattern, which as a rule is contaminated by random 
background hits, constitutes the input to the search for Cherenkov rings.
The ring search algorithm is described in Appendix 1.  
Precautions have been made in order to keep large Cherenkov rings.
In particular, the number of sectors to be treated with a given track	
was increased up to five (instead of 3 in the standard analyis), and number
of mirrors to be treated with a given track was increased up to four,
instead of one or two in the standard version.

The method of evaluation-of the probability that a given ring is 
due to background fluctuations, that 
fortuitously lead to the reconstruction of a ring image, is described in 
Appendix 2, the method being illustrated in the Figure in this Appendix and
by Fig.~\ref{fig:1}. Only the rings having the background probability below 
10\% were kept for further consideration.

In order to illustrate the above procedure several anomalous rings and their 
radial distributions are shown in Figs.~2-6.
\subsection{Background noise}
As mentioned above, apart from the photon detector hits that are due to
Cherenkov light produced by particles, traversing the liquid and gaseous
radiators, there are other (background) sources of hits
\cite{bloch,perez}:
\begin{itemize}
\item [a)] Electronic noise, induced by the LEP machine and other DELPHI
sub-detectors. This source produced several simultaneous hits on anode wires
and cathode strips.
\item [b)] Feedback photoelectrons. They are created during avalanche
development around the anode wires, when UV light emitted during the gas
amplification process liberates electrons in the presence of the TMAE.
Optical screens were mounted between the anode wires to minimize this effect,
confining it to the single wire from which the UV light was emitted.
However, sometimes feedback photoelectrons drifted also to neighbour wires.
\item [c)] Electronic oscillations, induced by large charges, 
mainly from ionizing tracks. These are characterized by several consecutive 
hits on the same channel separated by time intervals equal to the dead time. 
This gives additional secondary hits in the channel masking the signal hits.
\item [d)] Cross-talk between cathode strips. Large signals, mainly from
ionizing tracks, induced signals in neighbour strips of the same cathode
block, and occasionally in neighbour wires. This effect confused the coordinate
defined along wires for other real electrons coming just after.
%\item [e)] After-pulses, produced by the tail of large signal pulses
%superimposed on the background fluctuations. They appeared just after
%the dead time interval.
\item [e)] $\delta-$rays, created by the tracks crossing the drift tubes.
\item [f)] Ionization electrons ($dE/dx)$, produced by the passage of the
charged particles through the drift tubes and forming accumulations of
reconstructed points.
\item [g)] Cherenkov photons produced by particles when traversing the quartz
windows of the drift tubes.
\item [i)] Internal reflection (including total internal reflection): 
Cherenkov photons, produced in the liquid radiator and quartz windows, 
underwent internal reflection on the quartz boundaries. Some fraction of them  
might reach the photodetectors.
\item [j)] Hits from soft bremsstrahlung photons produced by electrons and
positrons in the DELPHI detector material.
\end{itemize}

These backgrounds were partially reduced by specific features in the detector
construction and by fine tuning of its working parameters (e.g., electronic
dead time, discriminator thresholds, MWPC settings). However they represent
the main source of spurious rings composed, mainly or in part, of hits of these
backgrounds.

\section{Event topologies studied}
\setcounter{equation}{0}
\renewcommand{\theequation}{3.\arabic{equation}}
In order to have the RICH patterns as clean as possible we restricted our
analysis to the simplest event topologies; in particular, hadronic events
were excluded from the consideration.\\ 
\vskip0.2mm
{\em Topology 1. Unlike-sign pair with a small oppening angle opposed by 
a high energy photon}. Schematically, this topology can be presented by the 
diagram given in Fig.~\ref{fig:6}a (the dashed lines in it represent the 
particles which go undetected). This topology corresponds to the reaction
\begin{equation} 
e^+ e^- \rightarrow \gamma ~x^+ x^- ,
\end{equation} 
where $\gamma$ denotes a high energy photon and $x^+, x^-$ stand for a pair of 
non-specified particles identified to be neither hadrons nor muons,
produced with a vanishingly small opening angle, 
the photon and the pair going in opposite directions.
In the case when the two charged particles are non-resolved by the TPC 
(topology 1a)
the common track $dE/dx$ is expected to correspond to 2~mips~\footnote{A mip 
is the $dE/dx$ of a minimum ionization particle.} or higher;
the case of resolved tracks corresponds to topology 1b.
Two examples of topology 1a and 1b events are given in 
Figs.~\ref{fig:61}-\ref{fig:161a}.\\
\vskip0.2mm
{\em Topology 2. Events with back-to-back or quasi-back-to-back tracks.}
The diagram of the back-to-back topology is presented in Fig.~\ref{fig:6}b. 
This configuration corresponds to Bhabha-like events, i.e.
\begin{equation}
e^+ e^- \rightarrow x^+ x^-, 
\end{equation}
called here topology 2a.
The diagram of a quasi-back-to-back topology (the final particles are 
acollinear but coplanar) is shown in Fig.~\ref{fig:6}c. 
This topology corresponds to $\gamma \gamma$ interaction events, i.e. 
\begin{equation}
e^+ e^- \rightarrow e^+ e^- ~x^+ x^-  
\end{equation}
(topology 2b).
Two respective examples of topology 2a and 2b events are shown in 
Figs.~\ref{fig:62a}-\ref{fig:162}.\\
\vskip0.2mm
{\em Topology 3. One-plus-three (2-jet) topology}. 
A search was also made for anomalous rings associated with tracks produced  
by secondary interactions of electrons
(or positrons), generated in the beam-beam or $\gamma \gamma$ interactions,
when these particles pass through the detector material, with the following
reaction taking place:
\begin{equation}
e X \rightarrow  X^\prime e~ x^+ x^-, 
\end{equation}
where $X$ and $X^\prime$ denote a charged object (a nucleus 
of the detector medium with which the incident electron interacts), 
before and after the interaction. Excepting the charged object, there are no 
undetectable charged particles in this reaction (see the reaction diagram in 
Fig.~\ref{fig:6}d). Thus a typical topology of events of this kind would be 
a presence of a ``jet" consisting of three tracks, of which at least two 
have non-zero impact parameters with respect to the primary vertex.
The particle producing the three-particle jet is expected to be opposed by 
another track in the opposite hemisphere, as shown in 
%the two examples of 
an example of this topology event in Figs.~\ref{fig:63},~\ref{fig:163}.
%,~\ref{fig:63a}.

\section{Selection of events having candidate anomalous rings}
The following selection cuts
%~\footnote{Selecting the events for our analysis, 
%we followed the commonly accepted practice in the searches for new phenomena 
%as well described in~\cite{vip} and cited below: 
%
%``Most of the particle collisions collected by the experiments' detectors
%are well-understood events, originating from known physics processes. When 
%it comes to searches for new phenomena, these events form a large amount of 
%background information, which has to be dealt with appropriately. The signal 
%could be a very tiny fraction of the selected events or, even worse, might end
%up in the fraction of events that are rejected by the trigger system.
%
%Given the current status of understanding and expectations regarding new 
%physics, it is not inconceivable that rare events might produce signatures
%in the detectors that are different from those commonly accepted. Thus,
%in this context, one could regard such events as anomalies. Semi-supervised
%and supervised algorithms ... could search for such anomalies and store them
%in a separate stream of the data, for off-line visual inspection."}  
were applied for the preliminary selection of 
events from the DELPHI DST's:

{\em The general selection of events} was done vetoing the hadronic events 
defined according to the DELPHI ``Team~4" criteria \cite{delphi2} \footnote
{The main criterion for selecting hadronic events was the requirement of
a multiplicity above 4 of charged particles with $p > 400$~MeV/$c$,
$20^\circ < \theta < 160^\circ$ and a track length of at least 30~cm 
in the TPC, with a total energy in these charged particles above 
$0.12 \times E_{cm}$.}. This retained about $10^7$~ events 
(called ``quasi-leptonic events"), found in the 1992 - 2000 data~\footnote
{The transition from LEP1 to LEP2 took place in the year 1995. 
LEP1 beam energies were centered around the mass of $Z^0$-boson
(91.2 GeV/$c^2$), the LEP2 period was aimed at physics beyond $Z^0$, with 
beam energies spanning between 161 and 208~GeV (the averaged energy being 
196~GeV).}. A further reduction of the amount of hadronic events was 
performed by using the DELPHI electromagnetic calorimeter as described below. 
Muons were identified (and rejected) with the HPC, HCAL and MUB responses. 
%As a next step of the selection, all charged tracks in 
%these events were required to be neither hadrons (as identified with the HPC 
%and HCAL responses) nor muons (as identified with the HPC, HCAL and MUB 
%responses). 
Another general selection criterion was the requirement that for all tracks 
of the events the errors of the momenta derived from the curvature of 
the tracks be  below certain limits 
%($\delta p/p < 0.05$ for low momentum tracks and 
($\delta p/p < 0.3$). 
%for tracks having momenta greater than 10~GeV/$c$). 
The tracks were furthermore required to be within 
the geometrical acceptance of the gas radiator of the DELPHI Barrel RICH 
($46^\circ < \Theta < 134^\circ$), with the barrel RICH being in 
an operational state. 

The further selection of events
used the cuts designed to enhance the contents of events of topologies 1, 2 
and~3 which are described below (the term $jet$ here corresponds to a single 
neutral or charged particle or two or three tightly bunched charged particles), 
and the results of the application of these selections to the DELPHI DST data 
is shown in Tables~1 and~2. 
\subsection{Topology 1}
\begin{itemize}
 \item [1.] Two jets in the opposite hemispheres are required in the event: 
one neutral jet and one jet consisting of 1 or 2 charged particles; each 
of the jets should have at least 50\% of the beam energy; in the case of the
neutral jet this energy should come from the HPC.  
 \item [2.] Track(s) of the charged jet should have associated shower(s) in
the HPC with the total jet shower energy exceeding half of the jet momentum,
the number of active HPC layers in the shower(s) being greater than or equal 
to 5 (of 9). 
 \item [3.] Track(s) of the charged jet should have the first measured point(s)
in the first layer of the DELPHI vertex detector (VD).   
 \item [4.] In the case of a single-track charged jet (topology 1a) the track
ionization has to be within the limits of $2.0 < dE/dx < 3.8$ mips
(corresponding to the 2 charged tracks being non-separable in the TPC),
with the number of TPC wires available for ionization measurement exceeding 40
out of a total of 192 wires (this cut aimed at the suppression of Compton event 
background, see Sect.~6.1).
 \item [5.] In the case of a jet with two separated tracks (topology 1b) 
the jet mass, calculated under hypothesis of the electron mass for the tracks, 
has to be below 0.5~GeV/$c^2$, and the smaller of the two track momenta 
has to exceed 4~GeV/$c$~\footnote{This cut was applied in order to keep the
efficiency of the track association with its HPC shower close to 100\%.}.
To suppress the Compton event background the jet acollinearity was required 
to be below $2^\circ$.  
\end{itemize}

It is worth to note that for a proper finding of Cherenkov rings (standard
or anomalous ones) only the track direction in the RICH, and not its 
momentum value, is required.
This direction is considered to be well defined even for the events in which
two tracks were not separated (not resolved) in the TPC, identified by  
double ionization only (the tracks from point~4 above). The reason for 
that these tracks were not resolved in the TPC was, 
in an addition to their negligible opening angle,
their high momenta leading to a small curvature of their trajectories
in the DELPHI detector magnetic field. This resulted in the appearance of
these tracks as practically straight lines in the detector; these tracks 
produced two Cherenkov rings in each of the RICH (liquid and gaseous) 
radiators (in the case of electron tracks all the rings are standard).
    
\subsection{Topology 2}
\begin{itemize}
 \item [1.] Two tracks of opposite charge are required in a 2-jet event, one
track per jet, both tracks having momenta greater than 4~GeV/$c$. 
Each jet should have at least 60\% of the beam energy (topology 2a, Bhabha-like
events) or, alternatively, each jet should have less than 60\% of the beam 
energy (topology 2b, $\gamma-\gamma$ events).
 \item [2.] Both tracks of the event should have associated showers in
the HPC with each shower energy exceeding half of the track momentum,
the number of active HPC layers in the showers being greater than or 
equal to 5 (of 9). 
 \item [3.] The tracks have to go in opposite directions in the $xy$ plane
having the track acollinearity lower than $2^\circ$ (in the case of 
topology~2a) or, in the case of topology 2b, having the track acollinearity 
higher than $2^\circ$ and the track acoplanarity below~$4^\circ$.
 \item [4.] Each of the tracks should have a gas Cherenkov angle,
%as calculated using the hits in the gas radiator of the Barrel RICH and 
stored on the DST, within the region 72 to 102~mrad~\footnote{The upper limit
is defined by a restriction on the gas Cherenkov radii imposed when producing
the DST's.}, with at least 5 photoelectrons associated with each ring;
 \item [5.] or, alternatively, one of the two tracks should have a
Cherenkov angle within the region 80 to 102~mrad,
with at least 5 associated photoelectrons.
% \item [4.] In the events of topology 2a each of the tracks should have an
%anomalous gas ring found with Barrel RICH hits by the computer-coded algorithm
%with the background probability less than 10\%, with the ring radius exceeding 
%72~mrad, the number of hits constituting the ring being greater than 
%or equal to~4. 
% \item [5.] In the events of topology 2b each of the tracks should have an
%anomalous gas ring found with Barrel RICH hits by the computer-coded algorithm
%with the background probability less than 10\%, with the ring radius exceeding 
%72~mrad, the number of hits constituting the ring being greater than or 
%equal to 4, or one such an anomalous ring in the event with the background 
%probability less than 1\%.
% \item [6.] No gas rings of radii from 57 to 67 mrad (standard rings) having 
%the background probability less than 10\% and containing more than 3 hits are 
%allowed in the event. 
\end{itemize}
%\subsection{Topology 3}
%\begin{itemize}
% \item [1.] There should be 1 or 2 charged particles in the event having 
%momenta measured in the TPC exceeding 4 GeV/$c$.
% \item [2.] At least one particle has to be identified as an electron with a
%shower in the HPC having at least two satellites.
% \item [3.] The electron track producing the shower with satellites must have
%a RICH response.
% \item [4.] The angular separation of the satellites from this track has to be
%less than $24^\circ$.
% \item [5.] The angular separation of the satellites from the nearest OD 
%track element has to be less than $12^\circ$.
%\end{itemize}
\subsection{Topology 3}
\begin{itemize}
 \item [1.] Two jets are required in the event: one jet consisting of a single
track, and the other consisting of 3 charged particles, with the overall sum
of the track charges equal to zero. All the tracks in the 3-particle jet should 
have momenta exceeding 3~GeV/$c$.
\item [2.] Tracks in the 3-particle jet should have associated showers in the 
HPC with each shower energy exceeding half of the track momentum, the number of 
active HPC layers in the showers being greater than or equal to 5 (of 9).
\item [3.] For a primary suppression of $\tau-\tau$ events (abundant in LEP1
data set) the effective mass of charged particles in the 3-particle jet
(calculated prescribing pion masses to the tracks) is demanded to be greater
than 0.5~$m_{\tau}$ (0.89~GeV/$c^2$), and the jet momentum is required to be 
greater than 0.8 of the beam momentum.
\item [4.] Two tracks in the 3-particle jet are required to 
have the first measured point at the detector radius $R > 35$~cm (i.e. outside
the ID, the DELPHI Inner Detector). And at least one track in this jet should 
have an impact parameter with respect to the primary vertex in the $xy$ plane 
exceeding 6~mm, or two tracks should have impact parameters each exceeding 4~mm,
while the impact parameters of all tracks with respect to the primary vertex 
should be below 10~cm.
 \item [5.] At least two tracks in the three-track jet should have a
non-standard RICH response from the gas radiator (stored on the DST), i.e.
either ring radii outside of the standard ring limits of $52 < r < 72$~mrad, 
or no response at all \footnote{This non-standard response was expected in the 
cases of anomalous rings having radii greater than 102~mrad.}, 
while each track is required to have some response from the liquid
radiator; these tracks should be within the angular acceptance of the gas
radiator i.e. within the range of $46^\circ < \Theta < 134^\circ$.
% \item [5.] Only one gas standard ring of radius from 57 to 67 mrad found by
%the computer-coded algorithm having the background probability less than 
%10\% and containing more than 3 hits is allowed in the 3-particle jet.
\end{itemize}

\subsection{Final selection}
The cuts described in Sects. 4.1 to 4.3 resulted in the selection 
of 395 events as primary signal candidate events of topologies 1, 2 and 3 
(which correspond to the sum of events in the Selection~4 row of Tables~1 
and 2 for LEP1 and LEP2 data, respectively).

A further analysis of all the selected events was done retrieving these events 
from the DELPHI raw data using the DELPHI event server, reconstructing 
the Barrel RICH hit patterns for all tracks in these events and searching for 
anomalous rings associated to each track, as described in Appendix 1. 

Of the 395 selected events 53~events were found to 
have at least one anomalous ring with a probability to be composed 
of background hits below 10\%; the distribution of these events
over different topologies is shown in Selection 5 rows of Tables 1 and 2. 
The minimal number of photons per ring was required to be 4.
In addition, the events of topologies~1 and~2 were required not to have 
standard rings in the gas radiator.

Of these 53 events 29 events were found to possess two gaseous anomalous rings 
(at least one anomalous ring per track in topology 2). All these events
were subjected to a further treatment, aimed at the finding of rings, 
standard and/or anomalous, in the liquid radiator. 
The search for liquid radiator rings in the events with the found gas rings, 
combined with the method of evaluation of the probability 
that the ring is composed of background hits, was performed 
as described in Appendix 2. The range of the liquid ring radii
searched for with this method spanned between 600 and 1100~mrad. 
In the case of such a ring being found with the ring radius exceeding 700~mrad
(excluding quartz ring range, see below)
it was considered as an anomalous liquid radiator ring; only those anomalous
liquid radiator rings having the background probability below 10\% 
were accepted for further analysis. 
Then the background probabilities of all anomalous rings in a given event 
were multiplied, and the product of the probabilities was required to be less 
than $10^{-3}$. Of the selected 29 events this left 9 events of topology 1, 
6 events of topology 2, and 12~events of topology 3, in total 27 events
as given in the Selection~6 rows in Tables 1 and~2. None of the rings in these 
events were compatible with being due to the systematic effects described 
in Section~5 below. The characteristics of the anomalous rings 
contained in all the 27 selected events are given in Tables 3 to 5. 
In total, the number of selected events with at least two anomalous rings
reduced to 27/17/11/6 when the combined probability that the reconstructed
rings are due to background, mimicking ring images, is below
$10^{-3}/10^{-4}/10^{-5}/10^{-6}$, respectively. 

These results can be considered as a primary evidence
for the presence of anomalous rings in the analysed data.

Note that topologies 1 and 2 contain only events coming from LEP2, 
i.e. no event of these topologies was found in LEP1 data. The significance 
of this observation will be discussed in Sect.~9.

\section{Systematic effects leading to the appearance of rings of radii
greater than standard ones}                    
There are several mechanisms which produced rings of radii greater than the
standard ring radii, in both radiators of the DELPHI Barrel RICH. They are well
understood and were readily identified when encountered in the data.

{\em Quartz rings.}
The liquid radiator of the DELPHI Barrel RICH has a quartz window of a thickness
of 4~mm which acts as an additional radiator. The refraction
index of the quartz has a dispersion, ranging from 1.527 to 1.627
within the effective photon wavelength band of the RICH. The corresponding
Cherenkov ring radii for ultra-relativistic particles
are between 857 and 909~mrad. 
%The mean number of
%photoelectrons in the quartz ring is expected to be 7. 
Due to the rather 
small angle of the total inner reflection at the boundary quartz-gas
(about $39^\circ$ as compared to $52^\circ$ for the analogous angle of the
liquid radiator) the quartz Cherenkov rings always appear as arcs and vanish 
at track polar angles greater than $79^\circ$ and below $101^\circ$. They are 
easily recognized by their specific radius and, in most cases, by a high
dispersion of the ring hits~\footnote{In several events the quartz radiator
also showed anomalous rings (indeed arcs), see e.g. 
Figs.~\ref{fig:13}, \ref{fig:14}. These rings (arcs) are quoted in Table~5, 
and after the recalculation of their radii to the liquid radiator ring radii 
using the proper relation for the refraction indices (1.577 for the quartz and 
1.273 for the liquid radiator) enter the correlation plot in Fig.~\ref{fig:15} 
below.}. 

{\em Jacobian rings.}
The principle of the RICH pattern recognition is the association of the 
Cherenkov photon emission point on a given track with a hit detected in the 
RICH photodetector by a light ray transported through the RICH media. When such
a ray, transported from the liquid radiator, enters the drift box at a small 
angle to the box surface this ray can often, in the case of a non-vanishing
background hit density in the corresponding region of the drift box, be 
associated with several different hits in the drift volume at small variations 
of the initial emission angle in the liquid radiator~\footnote{This happens 
when the ray incidence angle on the boundary {\em liquid-quartz} in 
the liquid radiator is close to the angle of the total internal reflection 
on a virtual boundary {\em liquid-gas}. Since all quartz window surfaces 
in a given sector are parallel, the refractions at the inner and outer surfaces 
of the windows cancel, and the resulting angle of the ray in a given drift box 
is the same as if the ray had passed a virtual boundary {\em liquid-gas} 
without a separating quartz window.}.
Mathematically, this means that a Jacobian between a volume element in the 
drift box to an angular element around a given ray in the liquid radiator is 
big. This results in the erroneous reconstruction of many rays with incidence 
angles slightly below the total inner reflection angle on the virtual boundary 
liquid-gas, which effectively means the appearance of apparently anomalous 
rings of radii of about 880~mrad. These ``Jacobian rings" are readily 
identified by their radii, by an unusually high hit density, by big Jacobian 
values and/or by big separations of the hit and the track $z$ coordinates
in the photodetector. Moreover, if the track polar angle is not close to 
$90^\circ$ (which is often  the case), Jacobian rings are shifted with respect 
to the predicted center.
 
{\em Standard liquid ring hits reconstructed as coming from the gas radiator.}
Due to statistical fluctuations in 
%the number of Cherenkov photons and fluctuation in their 
the absorption depth of the Cherenkov photons in the the drift gas,
the photons from the liquid radiator standard ring sometimes reach the outer
part of the drift box in a sufficiently big amount in order to create there 
a ring-like structure, especially at track polar angles close to $90^\circ$ 
(the probability for an individual photon from the liquid
radiator to reach the outer part of the drift box varied from 10\% to 30\% 
depending primarily on the photon trajectory angle). This structure 
can then be reconstructed by the gas radiator pattern recognition
as being produced by photons generated in the gas radiator and reflected by 
the focusing mirrors, resulting in an apparently anomalous gas radiator ring 
of a radius of $450 - 550$~mrad. Such rings are always shifted with respect to 
the predicted center, and can be identified by an observation of the ring of
a similar radius, reconstructed by the gas radiator pattern recognition over
the hits registered in the inner half of the drift box and having an 
essentially higher number of photoelectrons composing the ring.

{\em Variation of the gas radiator refraction index.}  
Variations of the pressure and temperature of the gas filling the detector
which affects the refraction index of the gas 
were under strict control during the RICH operation
since even tiny variations of the gas refraction index $n$ 
would worsen significantly the identification ability of the RICH
for ultra-relativistic particles (whose Cherenkov angles are close to
$cos^{-1} 1/n$, which is 62~mrad in our case). Therefore the refraction index 
of the DELHI gas radiator was kept within the limits of 
$1.00194 \pm 0.00001$~at Cherenkov photon energy 7~eV~\cite{rich4}.
For obtaining the Cherenkov angles exceeding 72~mrad from 
ultra-relativistic particles one has to
vary the refraction index to the values exceeding 1.00260. In particular,
in order to obtain such values as a result of an increase of the gas pressure 
it should be risen beyond 1380~hPa, while the working pressure in the DELPHI 
Barrel RICH was maintained at $1030 \pm 1$~hPa. Such an increase of the 
pressure value is outside of any realistic limit: a differential pressure 
of only 100~hPa would break the quartz windows \cite{delphi1}.
 
Thus this type of systematics can be excluded from the consideration.
  
\section{Studies of physical backgrounds}
Investigations were made under the hypothesis that events with anomalous 
rings are standard physics events which fit the non-Cherenkov-ring-related 
parts of selection criteria 1, 2 or 3 in Sects.~4.1-4.3, and in which 
the appearance of anomalous rings is due to
fortuitous combination of photons from several different and closely spaced
standard rings plus random background from hits produced by electrons and
positrons, showering in the DELPHI detector material, and from noise sources 
described in Sect.~2.2. Such events can be searched for
by modifying slightly the selection criteria described in Section~4 and 
analyzing them in a way similar to the events having 
anomalous ring event candidates in order to estimate 
the expected background level in an event sample of a given 
topology. Such an analysis, based on real data events, is described below. 

\subsection{Background studies of topology 1 events}
\subsubsection{Compton events}
\setcounter{equation}{0}
\renewcommand{\theequation}{6.\arabic{equation}}
In Compton scattering events a beam electron (or positron) scatters elastically
off a virtual photon emitted by the beam particle travelling in the opposite 
direction. The electron (or positron) that emits the photon usually escapes 
undetected into beam pipe, and the typical signature of a Compton scattering 
event consists in a photon, an acollinear (though coplanar) recoil electron 
(or positron) and nothing else,
thus mimicking the events of topology 1a. The momentum of the recoil electron
(or positron) $p_e$ and the directions of the two particles visible 
in the final state are tightly constrained by the relation \cite{compton}:
\begin{equation}
p_e=\sqrt{s} \Big{[}(1 + \cos\Theta_e) + 
\frac{\sin\Theta_e}{\sin\Theta_\gamma} (1 + \cos\Theta_\gamma) \Big{]}^{-1},
\end{equation}
where $\Theta$'s are the electron/positron and photon polar angles, 
respectively.

The selection of Compton scattering events for the cross-check analysis was 
done, from the whole LEP2 data set, with the same criteria as those listed in 
Sect.~4.1, with the requirement that the number of particles in the charged 
jet should be one and with the replacement of cut 4 by the requirement for 
the momentum of this particle to agree with the prediction by formula (6.1) 
within 20\%. We note that the requirement for both particles of this reaction 
to be within the barrel angular range and the energy cut of the selection 
criterion~1 (Sect.~4.1) eliminate the vast majority of the Compton scattering 
events constituting this type of background, resulting in a reduction of the 
sample of these events to 277~candidates, 140 of which were in the 20\% band 
around the formula~(6.1) predictions. However, after reinserting the cut~4 of 
Sect.~4.1 all these events have been removed, 
thus demonstrating that background from this process is negligible.
\subsubsection{Background of two-photon events}
Two photon events with one photon converted in the beam pipe
%(0.2\% of the radiation length) 
or in the 1st layer of the VD
%(0.3\% ??? of the radiation length) 
can imitate events of topology 1. 
This type of background can be studied by analyzing events
with visible photon conversions which occur outside the 1st layer of the VD,
but before the photons enter the TPC. 
%(the terms behind and in front here refer to the positions as seen from the 
%main vertex of a given event). 

The selection of events of this type for a cross-check of the analysis
was done using the same criteria as those listed in Sect.~4.1
with the replacement of cut 3 by the requirement that the first measured point
on the track(s) is outside the 1st layer of the VD. 225 events were found
in the whole LEP2 data with the radial distance of the first measured point 
distributed between 7.5~cm and 46~cm. All events were scanned looking for 
anomalous rings. 3 events were found containing anomalous rings satisfying 
all the conditions listed in Sect.~4.4.

By comparison, the number of topology 1 candidate events in LEP2 data, 
satisfying selection criteria 1-5 of Sect.~4.1, i.e. events having track(s) 
with the first measured point in the 1st layer of the VD, was found to be 25 
(in what follows, this subsample consisting of 25 events will be called ``the 
enriched sample of topology~1 candidates"). Among them there are 9~events 
containing two gaseous anomalous rings, satisfying the selection of Sect.~4.4. 
Having found 9 events among the 25 of this sample one expects 81 events
in the background sample of 225 events if they are of the same origin. 
Assuming that there are no correlations between the samples (this assumption
will be generally applied to all similar estimations below)
the probability to find 3 events in the background sample while expecting 81 
%is calculated, using the cumulative Poisson distribution, to be 
is negligible.
% (below $??? \times 10^{-34}$).

\subsubsection{Events of topology 1 after a tight selection}
Another estimation of the probability that the anomalous rings of events 
of topology 1 are rings fortuitously reconstructed from background hits
was done using the enriched sample of topology 1 candidate events, 
as defined above. As mentioned above, of the 25 events of this sample 9~events 
have been found to contain two gaseous anomalous rings and to satisfy all the
selection criteria for events of topology 1 described in Sects.~4.1 and 4.4,
including the absence of standard rings in the gaseous radiator.

Among the remaining 16 events 10 events were found to contain 1 or 2
standard rings in the gaseous radiator associated with the one- or 
two-particle jets and having no anomalous rings. These 10 events are therefore
identified as standard physics (background) events.
The remaining 6 events have no standard rings in the gas radiator, while 2~of
them have two anomalous rings per event, but the combined background probability
per event is higher than $10^{-3}$, two other events have single anomalous
rings, and the remaining 2 have no rings at all;
so these 6 events are ambiguous in the sense that 
their identification as anomalous ring candidates or as background remains 
undecided. Thus among the 25 events of the enriched sample there are 9 
topology~1 candidate events with two gaseous anomalous rings, 6 ambiguous 
events and 10~standard physics background events. The fraction of the 9 
topology 1 candidate events in the enriched sample, from which the identified 
background events are excluded (retaining 15~events), is $9/15 = 0.60$. 
If these 9~topology~1 candidates would be standard physics background events 
containing fortuitous anomalous rings, the expected number of such events 
among the identified standard physics events (10~events) is~6.0. 
Under the general assumption of no correlation between the samples the 
probability to find zero events containing two anomalous rings (i.e. topology~1
candidates) while 6 such events are expected is small, being 
below $2.5 \times 10^{-3}$.

{\em Conclusion of the sub-Section}.\\
Of these two independent estimations of the background levels the latter one 
(p-value less than $2.5 \times 10^{-3}$) is retained as being 
obtained with a more conservative approach to the background sample.

\subsection{Background studies of topology 2 events}
%\subsubsection{Dimuon events}
%The sample of dimuon events 
%(87,900 opposite-sign muon pairs selected from LEP1 data within 
%the Barrel RICH acceptance) has shown hit patterns with background hit 
%densities significantly lower than those of the patterns with electrons, 
%and therefore these events have been omitted from the background studies.   

\subsubsection{Topology 2a}
An estimation of the probability that a standard ring in a Standard Model 
Bhabha event would be mis-reconstructed as an anomalous ring was performed 
using a set of LEP2 events selected with the topology~2a cuts~1,~2 and~3 of 
Sect.~4.2. 
%A test of a background of the anomalous rings in events of topology 2a
%was performed using the Bhabha events.
%, as described in the reference \cite{test2}. 
A program-based search was made for an anomalous ring associated with 
one track in the event 
while the opposite track in the event was required to have a standard ring,
thus ensuring that a given event is a standard physics event and the apparently
anomalous ring found in it (if any) is a spurious one. The number of spurious 
anomalous rings in 3341 Bhabha events thus selected was found to be 27, the
fraction of such rings being equal to $27/6682 = (4.0 \pm 0.8) \times 10^{-3}$ 
per track. 

Having 24 candidate events of topology 2a in LEP2 data that have passed 
the preliminary selection as described in Sect.~4.2 (displayed on the Table~2 
row 4) containing 48~tracks the expected number of spurious anomalous rings in 
these events is calculated to be $48\times 4.0 \times 10^{-3} = 0.19$. One 
event of topology 2a (the event 110546:15356) containing two anomalous rings 
was found in this data set (see Table~4 and 
Figs.~\ref{fig:62a},~\ref{fig:162a}). The probability to find 
at least 2 (spurious) anomalous rings while the expected number of such rings 
is 0.19 equals to $ 1.6 \times 10^{-2}$  as calculated using the cumulative 
Poisson distribution~\footnote{The situation when two anomalous rings 
are found in the same event, decreasing the above probability, 
is considered in Sect.~8 below.}.  

\subsubsection{Topology 2b}
Two tests of a background of the anomalous rings in events of topology 2b
were performed.

The first estimation used minimally biased two-electron events, topologically
equivalent to those selected for the anomalous ring search 
in this topology candidate events. 
Relaxing the selection criteria described in Sect.~4.2, point~4 $-$ namely, 
requiring at least one gas anomalous ring of a radius exceeding 72~mrad with 
the minimal number of the hits on the ring to be 4, instead of requiring two 
rings of such a radius in the event with at least 5 hits on each ring, 
550~events were found in the LEP2 event sample; these 550 events contained 
all the 38 events of topology 2b quoted in the Selection~4 row of Table~2,
of which 5~events have passed the full set of selection criteria of Sect~4.4 
to be retained as good anomalous ring event candidates of a given topology. 
An equal number of 550 events was selected with the replacement 
of the requirement to have the DST information about 
possible anomalous rings by a requirement of having a photon with 
$E_\gamma > 20$~GeV. The latter condition lowered the track momenta and 
led to a track acollinearity, both features
being observed in anomalous ring candidate events of topology~2b. 
This sample was used to represent a background 
sample, unbiased with respect to the possible presence of anomalous rings
since the two samples did not contain common events.

No events of the background sample passed the full set of selection criteria 
for anomalous ring event candidates requiring two rings in the event, 
each of the two rings to have the probability less than 10\% 
of being fortuitous ring and the product of these probabilities 
to be less than $10^{-3}$, as described in Sect.~4.4 though five
events, each containing a single apparently anomalous ring, were found.
The purely statistical significance of the above analysis, resulting in zero
background sample events versus 5 observed events found in the work sample
is low (under the general assumption of no correlations between the samples 
it corresponds to a probability less than $6.7 \times 10^{-3}$).

The second estimation used the method of the background probability estimation
analogous to that described in Sect.~6.2.1.
The probability that a standard ring in a Standard Model $\gamma - \gamma$ 
event would be mis-reconstructed as an anomalous ring was estimated using a set 
of LEP2 $\gamma - \gamma$ events selected with the topology~2b cuts~1,~2 and~3 
of Sect.~4.2. A program-based search for an anomalous ring associated with 
one track in the event was carried out
while the opposite track in the event was required to have a standard ring, 
thus ensuring that a given event is a standard physics event 
and the apparently anomalous ring found in it (if any) is a spurious one. 
The number of spurious anomalous rings was found to be 62 in 819
$\gamma - \gamma$ events thus selected, the fraction of such rings being equal 
to $62/1638 = (37.9 \pm 4.8) \times 10^{-3}$ per track. 

Having 38 candidate events of this topology in LEP2 data which have passed 
the preliminary selection with topology~2b cuts as described in Sect.~4.2 
(displayed on the Table~2 row 4) containing 76~tracks, 
the expected number of spurious anomalous rings in 
these events is calculated to be $76 \times 37.9 \times 10^{-3} = 2.9$. 
Five events of this type, each containing two anomalous rings (see row 6 in 
Table~2), and four events, each containing a single anomalous ring, 
were found in this data set, i.e. 14 anomalous rings in a total. 
The probability to find at least 14 (spurious) anomalous rings while 
the expected number of such rings is 2.9 is small, being less than 
$2.4 \times 10^{-6}$ as calculated using the cumulative Poisson distribution.

For the further application the result of the first test of a background 
hypothesis for the topology~2b will be used as giving a more conservative 
estimation of the background probability value, i.e. 
less than $6.7 \times 10^{-3}$.\\
{\em Conclusion of the sub-Section}.\\
Assuming that there is no correlation between two subsamples of topology~2 
the combined probability value of the background hypothesis for this topology
is below $1.1 \times 10^{-4}$. 
 
\subsection{Background studies of topology 3 events}
The probability for the anomalous rings in topology 3 to be the result of
fortuitous combinations of background hits was estimated in a similar way
as that of the conservative approach described in Sect.~6.2.2.
The control sample of the standard physics events was constituted of events
selected under the same criteria as described in Sect.~4.3 with the replacement
of the requirement 5 of a non-standard RICH response in the gas radiator for
tracks in the three-track jets by the requirement for these tracks to have
at least 2 standard rings in this radiator. 145 such events were found in the 
data, containing no anomalous ring events satisfying our selection criteria.
This should be compared to 12 events satisfying these criteria which were
found in 125~events selected according to the complete set of the criteria
described in Sects.~4.3, 4.4 (8 events from LEP1 and 4 events from LEP2 data,
see Tables 1 and 2, last column). The probability to obtain
zero event yield from the control sample, while the expected number of such 
events is calculated to be $12 \times 145 /125 = 13.9$, is low (under general
assumption of no correlations between the samples the probability is below 
$ 1.0\times 10^{-6}$).

{\em Conclusion of the Section.}\\
It follows from the studies described in this Section that the probabilities 
that the events with the observed anomalous rings are due to fortuitous hit 
combinations in the topology 1, 2 and 3 event samples are low, being below 
$2.5 \times 10^{-3}$, $1.1 \times 10^{-4}$ and $1.0 \times 10^{-6}$, 
respectively, assuming no correlation between the samples. Under the 
assumption of no correlations and assuming further that there are no other 
systematic effects which we have overlooked, this results in a very low 
probability
%below $1.0 \times 10^{-9}$) 
that all the 3~topology event samples are composed entirely of events 
with fortuitous rings.

\section{An additional test of the background of fortuitous rings
(``night sky" method)}
%General study of the background of fortuitous rings
%using a method of a random shift of the Cherenkov plane origin}
An additional study of the background of fortuitous rings using a method of 
a random shift of the coordinate frame origin of the Cherenkov plane 
(called ``the night sky" analysis) was done and is described in this Section. 

\subsection{Description of the method}
In this study the hit patterns were used which have been associated with 
108 tracks (the choice of this amount will be explained below)
found in events selected with the criteria described in Sects.~4 and~6,
in which at least one candidate anomalous gas ring was found.

Since the goal of the present test was to
search for rings composed fortuitously from background hits produced by the
gaseous pattern recognition in the gaseous half of the drift boxes
only these ``gaseous" hits were treated by the test,
%by an automatic code, described in the Appendix~2, 
in the following way.

The predicted center of the Cherenkov ring associated with a given track 
(i.e. the origin of the coordinate frame of the Cherenkov plane for that track)
has been randomly displaced in this plane by a radial shift ranging, with 
the equal probability, from 30 to 200 mrad and (simultaneously) by a random 
azimuthal turn through an angle in the Cherenkov plane distributed uniformly 
from $0^\circ$ to $360^\circ$. Thus the original hit pattern lost the direct 
association with the corresponding track. Since the results resembled a night 
sky, this test was dubbed correspondingly.  

Each of the 108 aforementioned gaseous hit patterns, primarily  associated with 
a given track, was used to make the above displacements 50 times
thus giving 50 new ``track directions" providing in this way 5400 hit
patterns for spurious ring finding trials. This number was chosen
in order to greatly outnumber the number of 908 hit patterns produced
by tracks of 395 pre-selected events referred to in Sect.~4.4.    

The radial scan method described in the Appendix~2 was used to search for ring 
patterns. A ring was accepted if had 4 or more signal hits and the upper limit 
for the background probability was below 10\%.
This selection of the rings reproduces the analogous procedure
implemented for the selection of anomalous rings in the main analysis,
as described in Sect.~4. 
%and in the Appendix.

%In 108 hit patterns from the gaseous radiator with a non-displaced 
%origin of the Cherenkov plane scanned by the method of an automatic ring
%finding procedure, described above, there were 83 anomalous and 8 standard
%rings coming from the primary analysis, of which 63 rings were found by
%this procedure. Thus the efficiency of the method of the automatic ring finding
%for the DELPHI Barrel RICH data, with the requirement of a strong suppression
%of spurious rings as described above and in Sect.~4.4,
%can be evaluated as equal to $63/91 = (69\pm 5)$\%.

\subsection{Results of the method}
In 5,400 displaced hit patterns obtained from the original ones by 
random shifts of the origin of the coordinate frame of the Cherenkov plane 
and scanned by the method of an automatic ring finding procedure,
37 apparently anomalous rings were found, consisting of at least 4 hits and
satisfying the criterion to have an upper limit
for the probability to be combined of background hits of 10\%.
This corresponds to the probability of finding a spurious ring
by the above method in the DELPHI Barrel RICH background environment of
quasi-leptonic events of $37/5400 = 6.9\times 10^{-3}$.

To compare, 82 anomalous gas rings were found with the origin of the coordinate
frame of the Cherenkov plane not shifted coming from the 908 hit patterns 
mentioned in Sect.~7.1: 58 rings in the events with 2 gas anomalous rings per 
event and 24~rings in the events with single gas anomalous rings \footnote
{These numbers account for 27 signal events listed in Tables~3,~4 and~5, plus 2 
events of uncertain status having two anomalous rings mentioned in Sect.~6.1.3, 
while 24 events with single anomalous rings can be counted summing up the 
differences between the numbers in 5th and 6th rows of Tables 1 and 2 minus 
2 aforementioned uncertain events.}. 
The corresponding rate equals to $82/908 = 9.0 \times 10^{-2}$,
which is $13.0 \pm 2.6$ times higher than the ``night sky" background rate.

{\em Conclusion of the Section.}\\
The ``night sky" analysis background rate, scaled to the present analysis 
event sample statistics, predicts $37 \times 908/5400 = 6.2 \pm 1.0$ gaseous 
anomalous rings in this sample, versus 82 observed. 
The difference of $75.8 \pm 9.1$ rings 
%corresponds to 6.9 standard deviations. This excess 
is consistent with the sum of the 54 gaseous anomalous rings 
listed in Tables~3 to 5 (containing the final sample of 27 ``good" events) 
and the 24 single gaseous anomalous rings, found in the events 
satisfying the selection criteria given in Sect.~4.4. 

\section{Comparison of the rates of events with single 
and double anomalous rings}
An estimation of the probability of the hypothesis that the anomalous rings
observed in main analysis are fortuitous combinations of background hits can be
carried out comparing the rate of events with a single anomalous ring observed 
in the gaseous radiator with the rate of events containing two such rings
(an example of the event containing two gaseous anomalous rings, 
one ring per track, is presented in Figs.~2, 3). 
These two samples consisting of 29  and 24 events, respectively, 
contain, in total, 82~gaseous anomalous rings 
(each ring having the background probability less than 10\%) found 
in the initial sample of 395~events. Thus the average number of gaseous 
anomalous rings in these events is $82/395= 0.208 \pm 0.023$. 
Assuming that the rings are fortuitous combinations of hits and are independent
leads to the expected number of events in the sample containing
two such rings equal to $0.208^2 \times 395 = 17.0$. The probability to 
observe  29~events with two (gaseous) rings while the expected number of such 
events equals to 17.0 is, according to the cumulative Poisson distribution,
below $5.1 \times 10^{-3}$, which shows that the above assumption 
is unlikely to be valid.
 
%leading to a conclusion about the associated production of the anomalous rings.

This conclusion is applicable to any type of a background producing spurious 
rings when the processes underlying such a ring production
are independent, i.e. obeying Poisson statistics. 
This furthermore indicates a clear tendency for the observed anomalous rings 
to be produced in pairs.

\section{LEP2 versus LEP1}                                         
\setcounter{equation}{0}
\renewcommand{\theequation}{9.\arabic{equation}}
\subsection{Topology 1}
There are 9 candidates of topology 1 events in the final sample, listed in 
Table~3, all of them coming from LEP2 data. With the selection criteria applied
to candidates of topology 1 events, described in Sect~4.1, the principal 
background channel in this topology is the reaction
\begin{equation}
e^+ e^- \rightarrow \gamma \gamma
\end{equation} 
with one of the $\gamma$'s converted to $e^+ e^-$ pair in the beam pipe or 
in the 1st layer of the VD, with the subsequent production of at least two
fortuitous anomalous rings in the RICH from hits 
resulting  from showering electrons of the pair. 
Then one should expect that the number of events of this type in LEP1 data, 
satisfying all the selection criteria listed in Sect.~4 for events of this
topology, can be estimated using the ratio of the cross-sections 
of the reaction (9.1) at LEP1 \cite{glep1} and LEP2~\cite{glep2} 
which is $4.4 \pm 0.4$ \footnote
{This ratio is valid for the Barrel-region restricted cross-sections.},
multiplied by the ratio of the integrated luminosities corresponding to 
each data set (96~pb$^{-1}$ and 665~pb$^{-1}$, respectively), resulting in 
a factor 0.64. Thus, the expected number of events of this type of background 
is obtained to be 5.7. The probability to find zero events in LEP1 data,
while the expected number is 5.7 is low (the probability derived using Poisson
statistics is below $3.2 \times 10^{-3}$). This furthermore indicates that 
topology~1 events with anomalous rings seem to have an energy dependence 
of the production cross-section and, in turn, the production mechanism 
which differ from those of the reaction (9.1).
\subsection{Topology 2}
Among 6 events in the topology 2 final sample, listed in Table 4, all of them 
coming from LEP2 data, there is only one Bhabha-like (topology 2a) event, 
i.e. an event containing two collinear tracks. The track acollinearity in this
event equals to $0.1^\circ$ and its track momenta are measured to be 
$(100.5 \pm 6.5)$~GeV/$c$ and $(103.3 \pm 6.7)$~GeV/$c$, 
which agree well with the beam momentum of 103.0~GeV/$c$, 
thus corresponding to reaction~(3.2) (the event 110546:15356, shown in
Figs.~\ref{fig:62a}, \ref{fig:162a}). Assuming that the tracks 
in this event are electron tracks and the anomalous rings found in 
this event are fortuitous, one expects that the number of events with the 
same signatures (a high track collinearity and momenta agreeing with the beam 
one) in LEP1 data (coming from $Z^0$ decays to $e^+ e^-$) and containing two 
anomalous rings (one per track as the above event does), satisfying all the 
selection criteria listed in Sect.~4 for events of this topology, 
is 5.9 times higher. This number has been obtained 
as the ratio of the cross-sections of the reaction
\begin{equation}
e^+ e^- \rightarrow e^+ e^-
\end{equation}
at LEP1 \cite{elep1} and LEP2 \cite{elep2} reported for the
Barrel region, which is $40.8 \pm 4.5$, multiplied
by the ratio of the integrated luminosities corresponding to each data set 
(96~pb$^{-1}$ and 665~pb$^{-1}$, respectively).
The probability to find zero events in LEP1 data, while the expected number 
is 5.9 is low (the probability derived using Poisson statistics is 
$2.8 \times 10^{-3}$). This furthermore indicates that also topology 2 events 
with anomalous rings seem to have an energy dependence of the production
cross-section and, in turn, the production mechanism which differ from 
those underlying the reaction (9.2); we note however that this conclusion is 
only indicative being based on a single event of this type found at LEP2.

The comparison of LEP2 to LEP1 yields for topology 3 events was not tried
because of uncertainties in the required analysis.
%Such a comparison would require the convolution of electron (positron) spectra
%generated in $e^+ e^-$ interactions at various beam energies in the DELPHI 
%Barrel RICH acceptance with the interaction cross-sections of these particles 
%with the DELPHI detector material.
%, reduced by the ratio of of the corresponding luminosities. 
%The realization of an accurate study of these processes would require a lot
%of manpower and therefore was considered to be currently impossible for us.   

{\em Conclusion of the Section.}\\
When trying to find an explanation of anomalous rings within the framework of
the standard physics, one principal source resulting in the appearance of
spurious anomalous rings could be related to electrons (positrons) showering
in the DELPHI detector material, thereby producing background hits with 
possible mutual hit correlations. 
If so, when comparing LEP1 and LEP2 data, one would expect that the number
of standard physics events with spurious anomalous rings (in the samples of
events of topology~1 and~2) should be proportional
to the product of integrated luminosity times the relevant standard physics 
cross-section, which does not at all seem to be the case.

\section{Correlation between gaseous and liquid radiator rings}
\setcounter{equation}{0}
\renewcommand{\theequation}{10.\arabic{equation}}
An important feature of the DELPHI RICH was the presence
of two radiators: the liquid radiator and outside it, the gaseous one. 
With the use of the relationship originating from Cherenkov formula
\begin{equation}
  n_{liq} \cos{\theta_{liq}} = n_{gas} \cos{\theta_{gas}},
\end{equation}                         
the ring radius expected in the liquid radiator can be derived from the 
gaseous ring radius and {\em vice versa}. So, the question has been raised: 
whether the radii of the rings, found in the liquid and the gaseous radiators, 
obey formula (10.1)?

Fig.~\ref{fig:15} shows the anomalous ring radii observed in the gas radiator 
plotted against the ring radii observed for the same track 
in the liquid radiator for 27 events listed in Tables~3, 4 and~5 
(with some exceptions mentioned in the description of these Tables, 
see page~25), showing a clear correlation between the radii. 
The curved line in the figure is not a fit to the data, but has been derived 
using the relation (10.1) with $n_{liq} = 1.273$ and $n_{gas} = 1.00194$. 

The same correlation can be seen in Fig.~\ref{fig:16}, in which the variables 
on the axes have been transformed to $1/(n_{liq}\cos{\theta_{liq}})$ and  
$1/(n_{gas}\cos{\theta_{gas}})$, according to formula (10.1), which linearize 
the plot. The correlation coefficient calculated for 53 points of 
Fig.~\ref{fig:16} equals 0.992. The probability of obtaining such a coefficient
for uncorrelated variables was estimated on the basis of an 
analysis of $10^9$ MC ``toy experiments" in each of which the values of 
the 53 variable pairs were generated as distributed randomly according to the 
corresponding projections of the plot in Fig.~\ref{fig:16}. The probability 
resulting from this analysis is $10^{-9}$.  

The sum of $\chi^2$ for deviations of 53 points in Fig.\ref{fig:16} from the 
main diagonal is 40.1 The corresponding p-value for that these points are 
in agreement with the predictions of formula (10.1) is near 95\%.  
 
The accumulation of points in the left bottom corner of Fig.~\ref{fig:15} 
(a blow-up of this corner is shown in Fig.~\ref{fig:15a}) is explained by 
the fact that of the 53 gas rings, represented in Fig.~\ref{fig:15},
as many as 35 rings have a radius less than 120~mrad. 
In particular, all the events of topology~2 are concentrated 
in this corner owing to the selection cuts used, see Sect.~4.2 and Table~4: 
the anomalous gas rings with the radii below 120~mrad 
referred to in this Table 
correspond to the liquid radiator ring radii below 674~mrad, which are 
very close to the standard liquid ring radii $667 \pm 9$~mrad; 
a similar conclusion is applicable also for anomalous gas rings with radii 
below 240~mrad in the events of other topologies.

\section{Summary}
\begin{itemize}
\item [1).] The anomalous rings seen in the selected events,
all with the radii significantly exceeding the standard ones, are well centered
around the directions of the tracks with which they are associated.
\item [2).] All these tracks have the associated electromagnetic (HPC) showers 
with the shower energies agreeing with the track momenta~\footnote{In several 
events of topology 1 with two tracks non-resolved in the TPC the HPC shower 
energies exceed the measured track momenta, being however in a good agreement 
with the beam momenta.}, showing that the corresponding particles behave in 
the electromagnetic calorimeters like electrons. 
\item [3).] Each of the 27 selected events has at least two anomalous rings 
(several have more than two, up to four), with each ring having a probability 
less than 10\% of being reconstructed fortuitously from background hits, $and$
with a condition for the product of these probabilities (for all the rings 
in a given event, and assuming that the individual ring probabilities
are non-correlated) to be below $10^{-3}$. 
%In total, the number of
%selected events reduces as 27/17/11/6 when the product of the probabilities
%that the reconstructed rings are due to background, mimicking ring images,
%is below $10^{-3}/10^{-4}/10^{-5}/10^{-6}$, respectively.
\item [4).] The events selected {\em do not contain} standard gaseous rings 
unless such rings should be present in a given event (in events of topology~3).
On the other hand, if a given track produces an anomalous ring in the gaseous 
radiator with the radius smaller than 200 mrad, the expected close to standard 
ring associated with this track is seen in the liquid radiator.  
\item [5).] Careful studies of backgrounds based on data-driven tests 
(described in Sect.~6) result in the combined probability 
that the reconstructed rings are due to fortuitous combinations 
of background hits which is very low, 
%(as low as $1.0 \times 10^{-9}$ 
assuming that there are no correlation between the background samples.
\item [6).] An independent (``night sky") analysis of spurious rings composed 
of hits in the hit patterns, randomly shifted with respect to the expected 
impact positions of the corresponding detected tracks, gives an estimate 
for the number of fortuitous anomalous rings coming from background hits 
equal to $6.7 \pm 1.1$, while the actual observed number of anomalous rings 
associated with tracks is 82, the difference being $75.8 \pm 9.1$.
\item [7).] A comparison of rates of events with single and double gaseous
anomalous rings indicates a clear tendency for the observed anomalous rings
to be produced in pairs. 
\item [8).] A comparison of the number of anomalous rings observed in the
analysis of events of topologies 1 an 2 shows that there are significantly 
more anomalous rings found in LEP2 data than in LEP1 data which, 
after accounting for the cross-sections of the relevant background reaction 
(9.1) and (9.2), makes improbable the hypothesis that the events with anomalous
rings, presented in Tables~3 and~4, are fortuitously composed by hits from 
several standard rings produced by electrons (positrons) showering in 
the DELPHI detector material. This comparison also indicates that the 
production mechanisms of events of topologies 1 and 2, containing anomalous 
rings, are most likely to differ from the production mechanisms 
of the corresponding background reactions. 
\item [9).] In all the cases when anomalous rings associated with a given
track are observed in both the liquid and the gaseous radiators, the ring radii
are highly correlated in accordance with the respective refractive indices of 
these radiators (see formula (10.1) and Figs.~\ref{fig:15} and~\ref{fig:16}). 
The correlation coefficient for the points in Fig.~\ref{fig:16} is very high 
(0.992) and the probability to obtain such a correlation coefficient for 
uncorrelated data is very low.
%, estimated to be below $10^{-9}$.
\item [10).] At the time of the construction and subsequent operation of the
DELPHI detector computing capacity had not yet reached a level allowing a 
fully-fledged simulation program based on standard physics and a fully detailed
description of the detector.
%, and such a description cannot now be reproduced.
And there is not enough detailed information preserved about the DELPHI 
experiment to produce such a description today.  
Therefore, in the evaluation of probabilities quoted in this paper, it has not 
been possible to consider possible correlations between event samples 
analysed, implying that the validity of the evidence for the existence of 
anomalous rings that we have presented in this paper needs to be confirmed by 
other experiments.
\end{itemize}

\section{Conclusion}
The results of a search for events containing anomalous Cherenkov rings,
based on an analysis of the data of the DELPHI Barrel RICH,
are presented in this work.

A detailed study of backgrounds 
%and possible systematic effects, capable of producing apparently anomalous rings, 
indicates that the probability that the reconstructed rings are the result of 
fortuitous combinations of background hits is low. An important argument 
against the hypothesis that the observed anomalous rings are due to random 
combinations of background hits, is provided by 
the observation of a high degree of correlation between anomalous ring radii 
found in the liquid and gaseous radiators, obeying the Cherenkov cone formula. 

Although the results of the present analysis provide an interesting indication
of the existence of anomalous Cherenkov rings, 
%the evidence presented does not definitely establish the existence 
%of such rings. Further 
it is clear, that further searches for anomalous rings 
need to be made in future dedicated experiments 
in order to corroborate or refute the results of this analysis.
In particular, an analysis of events of $\gamma \gamma$ interactions in
Pb-Pb collisions at the LHC looks rather promising due to expected 
$Z^2$ enhancement factor in the cross-sections of these interactions 
which provide usually rather clean events. 

%For example, one can use for the anomalous ring search the existing data of 
%past and current experiments, e.g. the data of the ALICE experiment
%at LHC which contains a RICH (called HMPID)\cite{alice}. 
%%Events containing electron-like particles can be selected for the analysis; 
%The optimal processes for the investigation 
%could be $\gamma \gamma$ interactions (corresponding to
%our topology 2, reaction 3.3), with an expected $Z^2$ enhancement factor
%for the anomalous ring production in Pb-Pb collisions.
%The data of the LHCb experiment 
%%\cite{lhcbdet} 
%which contains RICH detectors with a high Cherenkov angle resolution 
%\cite{lhcb} 
%also can be used selecting 
%low multiplicity events in the data of pp, pPb and Pb-Pb collisions.

\subsection*{Acknowledgements}
\vskip 3 mm
We acknowledge with gratitude the effort provided by the DELPHI
Collaboration for the production of the  valuable experimental data used in
the present analysis.

We thank Profs. Yu~G.~Abov, K.~G.~Boreskov, F.~S.~Dzheparov, O.~V.~Kancheli
and Drs. D. Barberis, O.~N.~Ermakov, A.~I.~Golutvin, A.~A.~Grigoryan and 
S.~Sila-Novitsky for fruitful discussions and Drs. U.~Schwickerath, 
R.~M.~Shahoyan for technical assistance. 
%The help of DELPHI RICH experts
%M.~Dracos, P.~Kluit and D.~Liko was very important.
%Discussions with P.~Antilogus, B.~King, 
%M.~E.~Pol, J.~Timmermans, D.~Treille and 
%other members of the DELPHI Collaboration are greatly appreciated.

\newpage
%\vskip10cm
\section*{Appendix 1. Finding the rings by a computer-coded algorithm}
In order to find Cherenkov rings, composed of the hits in the hit patterns of 
the Barrel RICH gas radiator produced by the DELPHI pattern recognition from 
the data stored in the DELPHI raw data sets, at the stage of the event 
selection, a computer-coded algorithm was developed.
%, using an experience obtained during a visual scanning of Barrel RICH hit patterns.
Dubbed as 3-point method, it was used to search for standard and anomalous
rings in the events which have passed the preliminary selections based on
topological and kinematic characteristics of the events, described in 
Sect.~4.1 to Sect.~4.3. 
% of both data sets~\footnote{
%As mentioned earlier (see Sects.~2.1 and~4.2) the hits which would correspond
%to gas ring radii exceeding 102~mrad were not stored in the DST data,
%but were available in the raw data.}.
  
The principle of the algorithm
%, aimed at the automatic search for the rings, 
was to try all the combinations of 3~hits, present in the hit pattern 
associated with a given track (hence the name ``the 3-point method")
and calculate the parameters of the ring going through these 3 points:
the ring radius and the two coordinates of the ring center on
the Cherenkov plane (for the definition of the plane see Sect.~2.1).

The ring center coordinates $x_0$ and $y_0$ were allowed to deviate from
the plane frame origin by 10~mrad in the $y$ direction (i.e. in the
magnetic-field bending plane) and by 5~mrad in the $x$ direction
(perpendicular to the magnetic-field bending plane) in the case of the
potential ring radius below 110~mrad. 
%(in both the DST and the raw data sets).
When searching for rings 
%in the raw data the rings 
of larger radii (up to 1000~mrad), the tolerances for the ring center 
deviations were increased proportionally to the ring radii.

For each ring selected in this way a search for additional hits, lying
within $\pm 1 \sigma_{p.e}$ of the individual photo-electron error,
was carried out, the minimal number of hits lying on the ring being
required to be 4. Of the many ring candidates reconstructed from 3-hit
combinations those giving the minimal probability for the ring
to be composed of background hits, calculated with the Poisson formula as
described in Appendix~2, were retained, both standard and anomalous rings.
In order to accept a ring (of both types) these probabilities were required to 
be below 10\%. The information about the rings found, standard and anomalous, 
was used 
%first in the preliminary selection of events in the DST data, 
%described in Sects.~4.1 - 4.3, and then 
in the final selection of events as described in Sect.~4.4.

\vskip1cm
\section*{Appendix 2. The method of estimating the probability
that an anomalous ring can be composed of RICH background hits}
The raw data RICH hit patterns produced by tracks in the selected events were 
scanned using an algorithm with a sliding radial window. 
%of width roughly equal to two times the expected single photon angular 
%standard error~\footnote{
The full widths of the sliding windows
were fixed to be 8~mrad for gas ring radii below 100~mrad, 12~mrad for gas ring 
radii between 100 and 200~mrad, 24~mrad for gas ring radii between 200 and 
400~mrad and 48~mrad for gas ring radii exceeding 400~mrad.
For liquid rings the full width of the sliding window was 48 mrad in the whole 
liquid ring radii range studied which was spanned between 600 and 1100~mrad.
%}. 
The scanning step used 
was equal to 1/8 of the window width. The number of hits inside 
the window was counted at each step of the scanning procedure (hits 
overlapping within the individual photon angular error were counted as a 
single hit) and fed into a histogram, called the raw radial distribution, 
the histogram bin width being equal to the scanning step. The maxima in this
raw radial distribution indicate the most probable Cherenkov cone
angles. In parallel, another histogram, called weighted radial distribution, 
was filled. The content of each bin of the raw radial distribution was 
transformed to the probability of the hypothesis that the content of this
bin was a result of an accumulation of background hits. Then the negative
decimal logarithm of this probability was plotted for each radial bin.
The probability was calculated assuming a Poisson distribution for the 
background hits inside the bin, the mean level of the background 
being estimated from the hit density in two adjacent annuli~\footnote{An area 
between two concentric rings will in this paper be called an ``annulus".}
inside and outside of the scanning window (called background annuli), as shown 
in the following figure:
\vskip1cm
\includegraphics[height=15.0cm,width=8.0cm]{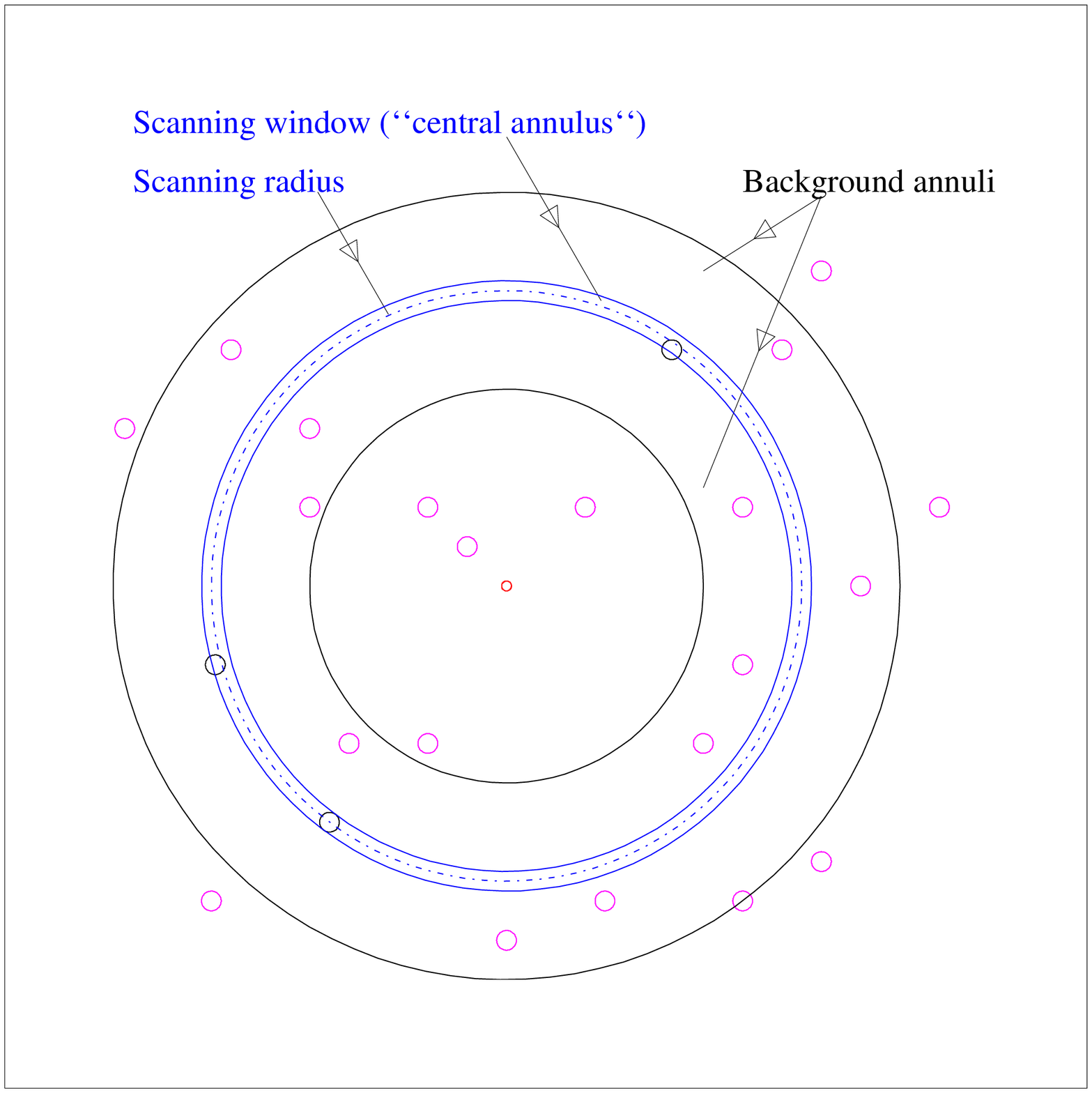}

\begin{figure}
\begin{center}
\epsfig{file=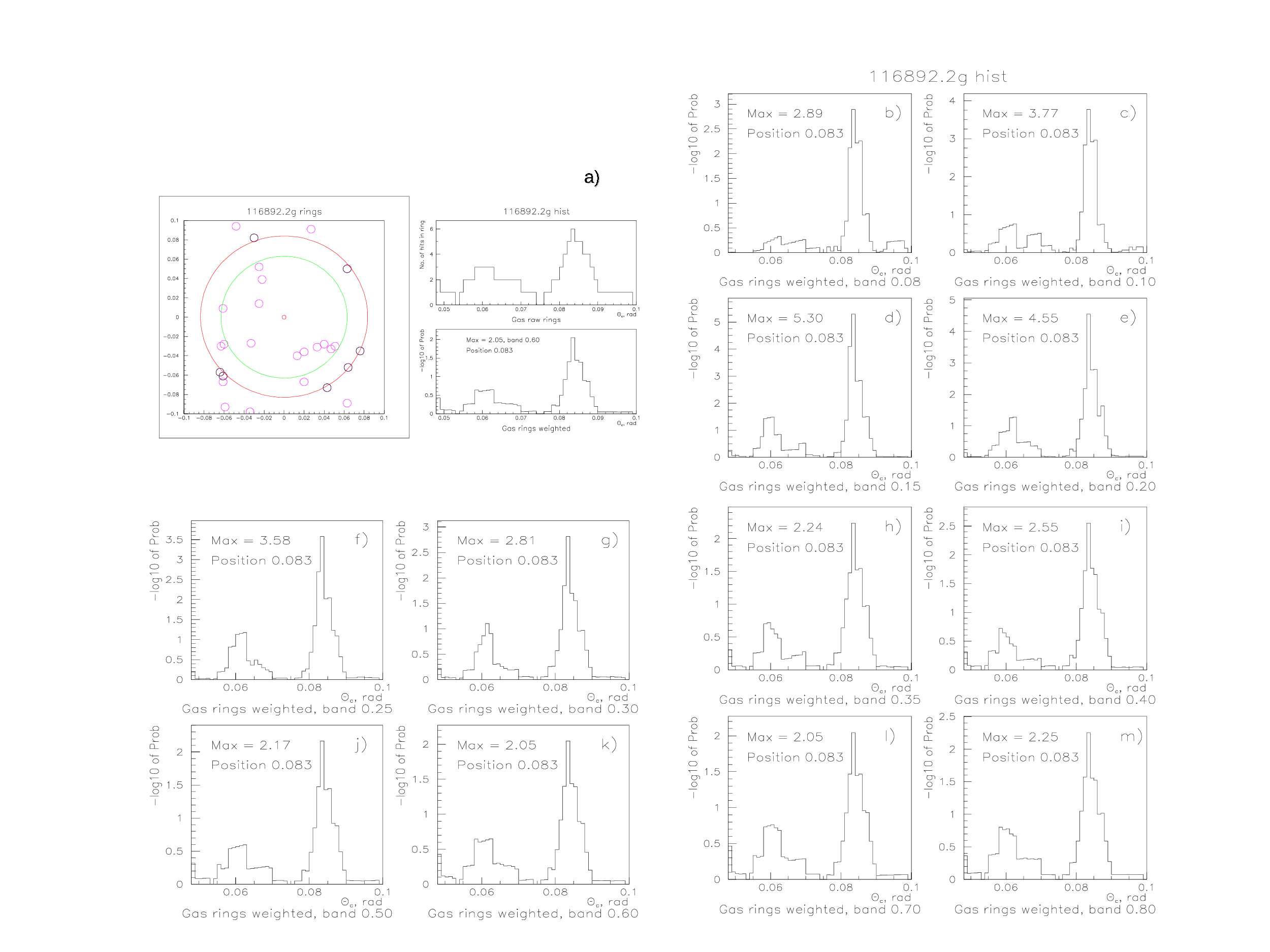,bbllx=100pt,bblly=0pt,bburx=650pt,bbury=570pt,%
width=17cm,angle=0}
\end{center}
\caption{Illustration to the the procedure of finding the maximum probability
for an anomalous ring to be composed of background hits. Panel {\bf a)} shows
an anomalous ring of a radius of 83~mrad and its radial distributions selected
to represent a given anomalous ring (see below); the inner green circle in this
panel marks the position and size of a standard ring (not seen in the pattern). 
Panels from {\bf b)} to {\bf m)} show 12 weighted radial distributions obtained
with different background annuli widths (see the text, Sect.~2.1 and 
Appendix~2). The maximal background probability found with the width of the 
background annulus equal to 0.60 of the scanning radius (panel {\bf k)}); 
this plot is that which is shown in the right lower part of panel {\bf a)}. }
\label{fig:1}
\end{figure}

\vskip4mm
Since the background hit density was not uniformly distributed, 
the widths of the background annuli were varied  
in 12 steps from 8\% to 80\% of the scanning radius. 
For $n$ hits found in a given scanning window (i.e. in the central annulus)
the probability of the number of background hits in this annulus $\geq n$ was
calculated using the cumulative Poisson distribution.
After these calculations had been performed for all background annuli
widths, the width giving the maximal probability for the anomalous ring
under consideration to be composed of background hits was selected
to represent the weighted radial distribution,
as illustrated in Fig. \ref{fig:1}a.
This maximal probability 
%corresponding to the minimal height of the relevant
%peak in the 12 radial distributions weighted in such a manner)
is quoted, in this analysis, as the upper limit for the
background probability for a given ring, only those anomalous rings being
accepted which had this probability below 10\%. The peak position 
in the corresponding distribution determined the Cherenkov angle.   
%, overriding the values obtained with the 3-point method (or predicted 
%from the gas ring radius in the case of a liquid radiator ring, see Sect.~10).

%\newpage

\newpage
\onecolumn
{\bf Description of the contents of Tables 1 and 2 below.}\\ 
Topologies 1 and 2 are subdivided in two sub-topologies, $a$ and $b$, as 
defined in the table heads. Selection~1 means the general selection and primary
topological cuts (point 1 in the list of selection cuts, Sects.~4.1-4.3).
Selection~2 retains the tracks having electron-like nature 
(point 2 in the list of selection cuts);
thus these two selections precede the application of special cuts which depend 
on a particular topology. Selection~3 means final topological and kinematic 
cuts preceding the application of cuts based on the Barrel RICH information: 
for topologies 1a, 1b this means cuts 3, 4 and 5 specified in Sect.~4.1; 
for topologies 2a, 2b this means the cut 3 specified in Sect.~4.2; and the 
cuts~3 and~4 for topology 3 (Sect.~4.3). Selection~4 means the application of 
cuts based on the Barrel RICH information stored in the DST's, these cuts being
specified: for topologies 1a, 1b: not applied; 
for topologies~2a and~2b the cuts in points~4 and 5 of Sect.~4.2; 
for topology 3 the cut in point~5 of Sect.~4.3. 
%Selection~5 lists the number of events retained after visual scanning of the 
%pre-selected hit patterns. 
Selection 5 gives the number of events passed the 3-point method processing
with the use of the RICH information from the raw data,
found to contain at least one anomalous ring with the upper limit for the 
background probability less than 10\%, 
%(i.e. the events processed by the automatic code mentioned in Sect.~4.4) 
and which have no standard gas rings in the events of topologies~1 and~2. 
Selection~6 means the final selection which requires the presence of at least 
two anomalous rings in each event with the combined background probability
below $10^{-3}$; the numbers in this row 
present the events which are stored in Tables 3 to 5.    

\vskip0.5cm

{\bf Table 1.} Selection of LEP 1 events: the number of events retained after
an application of sequential cuts described in the comments to the Tables~1,~2 
above and in the text, Sect.~4.
\begin{center}
\begin{tabular}{ |c |c |c |c |c |c |}
\hline
          &&&&&\\
Selection&   Topology 1a,    &  Topology 1b, & Topology 2a, & Topology 2b,  & Topology 3 \\
         &non-resolved tracks&resolved tracks& Bhabha-like  &$\gamma-\gamma$&            \\
          &&&&&\\
\hline
               &&&&&\\
   1     &     13897         &     522       &  206089      &   26985       &    9528    \\
               &&&&&\\
   2     &      7134         &     198       &   69223      &    2352       &    1050    \\
               &&&&&\\
   3     &       22          &      43       &   47771      &     268       &     247    \\
               &&&&&\\
   4     &       22          &      43       &     87       &      30       &     117    \\
               &&&&&\\
%   5     &        6          &      12       &     23       &      10       &      36    \\
%               &&&&&\\
   5     &        3          &       1       &      7       &       0       &      13    \\
               &&&&&\\
{\bf 6}  &   {\bf 0}         &    {\bf 0}    &   {\bf 0}    &   {\bf 0}     &  {\bf 8}   \\
          &&&&&\\
\hline
\end{tabular}
\end{center}
%\vskip0.5cm
\newpage

{\bf Table 2.} Selection of LEP 2 events: the number of events retained after
an application of sequential cuts (see comments to Tables~1 and~2 above 
and the text, Sect.~4).
\begin{center}
\begin{tabular}{ |c |c |c |c |c |c |}
\hline
          &&&&&\\
Selection&   Topology 1a,    &  Topology 1b, & Topology 2a, & Topology 2b,  & Topology 3 \\
         &non-resolved tracks&resolved tracks& Bhabha-like  &$\gamma-\gamma$&            \\
          &&&&&\\
\hline
               &&&&&\\
   1     &     2302          &     131       &  10644       &   19589       &     214    \\
               &&&&&\\
   2     &     1975          &      82       &   5889       &    3050       &      42    \\
               &&&&&\\
   3     &      19           &       7       &   4658       &    1752       &      10    \\
               &&&&&\\
   4     &      19           &       7       &    24        &     38        &       8    \\
               &&&&&\\
%   5     &      16           &       4       &     9        &     13        &       8    \\
%               &&&&&\\
   5     &       9           &       4       &     1        &      9        &       6    \\
               &&&&&\\
 {\bf 6} &  {\bf 6}          & {\bf  3}      &{\bf 1}       & {\bf 5}       &  {\bf 4}   \\
               &&&&&\\
\hline
\end{tabular}
\end{center}
\vskip1.0cm
{\bf Description of the contents of Tables 3-5 below.}\\ 
These Tables contain only events in which there are for each track both a gas
and a liquid ring (with two exceptions described below). 
If in the ``Ring radius" column there is a liquid ring, 
marked ``Liq", preceded by a gas ring, marked ``Gas", this means that the 
liquid ring is anomalous and pertains to the same track as the gas 
ring. If a gas ring is not accompanied by a liquid ring this means that 
the liquid ring pertaining to the same track as the gas ring has a radius 
$< 700$~mrad, in which case the liquid ring parameters have been omitted from 
the Tables. However, both the gas and the liquid rings for each track have been
plotted in Figs.~\ref{fig:15} and~\ref{fig:16} below (there are two exceptional 
events: the event 107038:11748 of topology~1a in which the expected liquid 
radiator ring, with the radius of 854~mrad predicted from the gaseous anomalous
ring radius, overlaps strongly with the quartz ring making its unambiguous 
identification impossible, and the event 49062:2354 of topology 3 in which a 
track producing an anomalous ring of a radius of 220~mrad in the gas radiator
went through the crack in the liquid radiator). 
The OD tag indicates the state of the association of the Outer
Detector track element with the track to which a given gas ring was assigned:
2 means a clear association, 1 means that a cluster of OD hits exists in the
OD region intersected by the track, which prevents establishing a clear 
association, and 0 means that no association was found. The numbers in the 
4th column of the Tables indicate the number of RICH hits found in the radial 
window of a width of two times of the expected single photon angular accuracy, 
corresponding to the ring radius. The numbers in the 5th column of the Tables
present the expectations for the background hit levels in these windows
calculated on the base of the hit density in the vicinity of the ring, as
explained in Appendix~2. The numbers in the 6th column present the upper 
limits for the probability of the ring to be composed of background hits.

\newpage
{\bf Table 3.} Characteristics of anomalous rings found in events of topology 1.
\begin{center}
\begin{tabular}{ |c |c |c |c |c |c |c |}
\hline
%          &&&&&&\\
 Run  &Ring radius,   &OD tag&No. of hits& Expected & BG p-value  &  Product of BG      \\
Event &   mrad        &      &found  & BG level  &  per ring       & p-values per event \\
\hline
          &&&&&&\\
71768 &Gas~$347\pm~15$&  1   &   6   &   2.72    &     0.0587       &$1.0\times 10^{-5}$\\
6277  &Liq~~$741\pm19$&      &  10   &   3.84    &     0.0062       &                   \\
      &Gas~~$~85\pm~3$&  1   &   6   &   2.26    &     0.0280       &                   \\
%          &&&&&&\\
%83700 &Liq  $700\pm18$&    &  14     &   6.22    &$4.9\times10^{-3}$&$2.5\times 10^{-5}$\\
% 574  &Gas~~$~90\pm~4$&    &   5     &   2.38    &     0.0933       &                   \\
%      &Gas  $265\pm11$&    &   5     &   2.01    &     0.0537       &                   \\
          &&&&&&\\
{\em84451}&Gas $460\pm16$&  2   &   16  &  10.46    &     0.0667    &$3.0\times 10^{-4}$\\
{\em 478}&Liq  $786\pm20$&      &    7  &   2.69    &     0.0560    &                   \\
      &Gas  $102\pm~5$&  2   &    4  &   1.61    &     0.0801       &                   \\
          &&&&&&\\
%85618 &Liq~~$722\pm19$&      &    7  &   2.37    &     0.0110       &$5.6\times 10^{-5}$\\
%21487 &Gas~~$~80\pm~3$&      &    4  &   1.58    &     0.0759       &                   \\
%      &Gas  $342\pm13$&      &    4  &   1.52    &     0.0676       &                   \\
%          &&&&&&\\
{\em86748}&Gas $335\pm14$&  2   &    5  &   0.90    &     0.0024    &$1.0\times 10^{-6}$\\
{\em 757}&Liq~~$747\pm19$&      &   11  &   6.08    &     0.0459    &                   \\
      &Gas~~$~79\pm~3$&  2   &    5  &   1.26    &     0.0094       &                   \\
          &&&&&&\\
%88024 &Gas~~$~80\pm~3$&  2   &    8  &   2.95    &     0.0109       &$<1.0\times10^{-7}$\\
% 1284 &Gas  $298\pm11$&  2   &   11  &   0.36    &   $<0.0001$      &                   \\
%          &&&&&&\\
101830&Gas~~$~75\pm~3$&  1   &    9  &   2.77    &     0.0022       &$7.0\times 10^{-5}$\\
 15719&Gas  $107\pm~4$&  1   &    9  &   4.29    &     0.0315       &                   \\
          &&&&&&\\
102500&Gas  $816\pm34$&  0   &    5  &   1.74    &     0.0320       &$3.8\times 10^{-6}$\\
 9258 &Liq~$1010\pm27$&      &    9  &   2.79    &     0.0024       &                   \\
      &Gas~~$~85\pm~3$&  0   &    5  &   1.96    &     0.0492       &                   \\
          &&&&&&\\
{\em106061}&Gas~$885\pm35$&  2   &    9  &   3.57    &     0.0111    &$1.1\times 10^{-6}$\\
{\em3293}&Liq~$1052\pm23$&      &    7  &   2.62    &     0.0180    &                   \\
      &Gas~~$~91\pm~4$&  2   &    7  &   2.06    &     0.0053       &                   \\
          &&&&&&\\
107038&Gas~~$~87\pm~4$&  1   &    7  &   2.41    &     0.0117       &$4.7\times 10^{-4}$\\
 11748&Gas~$583\pm~22$&  1   &    7  &   3.07    &     0.0369       &                   \\
          &&&&&&\\
116892&Gas~$~266\pm12$&  1   &    4  &   1.44    &     0.0578       &$8.1\times 10^{-6}$\\
 5928a&Liq~~$702\pm17$&      &   10  &   4.43    &     0.0155       &                   \\
      &Gas~~$~83\pm~3$&  1   &    6  &   1.74    &     0.0090       &                   \\
          &&&&&&\\
116892&Gas  $831\pm30$&  2   &   10  &   4.90    &     0.0283       &$7.9\times 10^{-8}$\\
 5228b&Liq~$1021\pm27$&      &    5  &   1.67    &     0.0279       &                   \\
      &Gas~~$~75\pm~2$&  2   &   14  &   4.14    &     0.0001       &                   \\
%          &&&&&&\\
%117471&               &       &     &           &                  &                   \\
% 1512 &               &       &     &           &                  &                   \\
%          &&&&&&\\
%117637&               &       &     &           &                  &                   \\
% 4545 &               &       &     &           &                  &                   \\
%          &&&&&&\\
\hline
\end{tabular}
\end{center}
%\vskip0.4cm
Comment to Table 3. Events having event numbers printed in italic pertain to 
topology~1b; the rest events are of topology 1a. 
Comments to Tables 3, 4 and 5: Multiplying the background p-values of the
individual anomalous rings in an event to obtain a combined p-value requires 
that the anomalous rings found in a given event are uncorrelated. We have found
it difficult to evaluate the correlations of the rings, but assume that they
are very small. In consequence, the values of the products of p-values in the
rightmost columns of these Tables provide only indicative upper limits for the
probability of that all of the anomalous rings in each event are due to
fortuitous background combinations.  

\newpage
\vskip1.0cm
{\bf Table 4.} Characteristics of anomalous rings found in events of topology 2.
\begin{center}
\begin{tabular}{ |c |c |c |c |c |c |c |}
\hline
          &&&&&&\\
 Run  &Ring radius,   &OD tag&No. of hits&  Expected & BG p-value     & Product of BG   \\
Event &   mrad        &      &   found  &  BG level  &  per ring      &p-values per event\\
\hline
          &&&&&&\\
%76039 &Gas~~$~81\pm~4$&  2   &    6     &   1.98    &     0.0160       &$6.7\times 10^{-4}$\\
% 435  &Gas~~$~77\pm~3$&  2   &    5     &   1.87    &     0.0417       &                   \\
%          &&&&&&\\
77190 &Gas~~$~84\pm~3$&  2   &    7     &   1.52    &     0.0010       &$5.9\times 10^{-5}$\\
 2950 &Gas~~$~86\pm~4$&  2   &    4     &   1.45    &     0.0589       &                   \\
          &&&&&&\\
%80957 &Gas~~$~80\pm~3$&      &    6     &   1.94    &     0.0144       &$7.3\times 10^{-4}$\\
%10048 &Gas~~$~72\pm~3$&      &    4     &   1.37    &     0.0508       &                   \\
%          &&&&&&\\
%83450 &Gas~~$~89\pm~4$&      &    5     &   1.28    &$9.9\times10^{-3}$&$5.1\times 10^{-4}$\\
% 1745 &Gas~~$~88\pm~4$&      &    6     &   2.63    &     0.0511       &                   \\
%          &&&&&&\\
85371 &Gas~~$118\pm~5$&  0   &    8     &   3.16    &     0.0158       &$1.6\times 10^{-6}$\\
  626 &Gas~~$104\pm~4$&  2   &    7     &   0.96    &     0.0001       &                   \\
          &&&&&&\\
88975 & Gas~~$~72\pm~3$& 0 &    5      &   1.23    &     0.0086       &$7.7\times 10^{-4}$\\
 84   & Gas~~$~72\pm~3$& 2 &    4      &   1.67    &     0.0895       &                   \\
          &&&&&&\\
105033&Gas~~$~76\pm~2$&  2   &    7     &   1.06    &     0.0001       &$7.5\times 10^{-7}$\\
 3999 &Gas~~$~78\pm~4$&  0   &    7     &   2.13    &     0.0064        &                   \\
          &&&&&&\\
{\em110546}&Gas~~$~75\pm~3$&  2   &    4&   0.92    &     0.0147       &$6.9\times 10^{-4}$\\
{\em15356}&Gas~~$~75\pm~4$ &  2   &    4&   1.34    &     0.0470       &                   \\
          &&&&&&\\
115187&Gas~~$~75\pm~3$&  1   &    5     &   1.65    &     0.0268       &$5.7\times 10^{-4}$\\
 4860 &Gas~~$~78\pm~4$&  1   &    4     &   1.04    &     0.0215       &                   \\
          &&&&&&\\
\hline
\end{tabular}
\end{center}
\vskip0.6cm
\noindent Comments to Table 4. The event 110546:15356, having event number 
printed in italic, pertains to topology 2a; the rest events are of topology 2b.
The absence of liquid rings in this table is explained by the fact that the
liquid rings seen associated with the gas rings all have radius less than
674~mrad , very close to the standard liquid ring radius 667~mrad.
%All the rings presented in this Table have 
%corresponding standard (or almost standard) rings of a radius of $< 700$~mrad
%in the liquid radiator. The OD tag meanings are the same as explained in
%comments to Table 3. The numbers in the 4th column of the
%Table indicate the number of RICH hits found in the radial window of a width
%of two times of the expected single photon angular accuracy, corresponding
%to the ring radius. The numbers in the 5th column of the Table
%present the expectations for the background hit levels in these windows
%calculated on the base of the hit density in the vicinity of the ring.
%The numbers in the 6th column of the Table present the upper limits for the
%probability of a given ring to be composed of background hits.
\newpage
{\bf Table 5.} Characteristics of anomalous rings found in events of topology 3.
\begin{center}
\begin{tabular}{ |c |c |c |c |c |c |c |}
\hline
          &&&&&&\\
 Run  &Ring radius,   &OD tag&No. of hits&  Expected & BG p-value      &Product of BG     \\
Event &   mrad        &      &  found    &  BG level &  per ring       &p-values per event\\
\hline
          &&&&&&\\
41633 &Gas~~~$72\pm 3$&  2   &    6      &   2.22    &      0.0260      &$3.4\times 10^{-4}$\\
 1568 &Gas~~~$89\pm 5$&  2   &    7      &   2.45    &      0.0129      &                   \\
          &&&&&&\\
42495 &Gas~~~$82\pm 4$&  2   &    8      &   4.10    &      0.0570      &$8.9\times 10^{-4}$\\
24853 &Gas~~$107\pm 5$&  2   &   11      &   5.10    &      0.0156      &                   \\
          &&&&&&\\
46877 &Gas~~$815\pm32$&  2   &    9      &   3.77    &      0.0002      &$2.4\times 10^{-7}$\\
 4278 &Liq~~$987\pm26$&      &    6      &   3.06    &      0.0897      &                   \\
      &Gas~~$181\pm~7$&  2   &   12      &   6.25    &      0.0263      &                   \\
          &&&&&&\\
49062 &Gas~~~$84\pm~4$&  2   &    8      &   2.34    &      0.0028      &$9.4\times 10^{-5}$\\
 2354 &Gas~~$220\pm10$&  0   &   10      &   4.35    &      0.0336      &                   \\
          &&&&&&\\
49316 &Gas~~~$76\pm~4$&  2   &    5      &   1.89    &      0.0430      &$6.3\times 10^{-4}$\\
 584  &Gas~~$183\pm 9$&  2   &    7      &   2.51    &      0.0146      &                   \\
          &&&&&&\\
50971 &Gas~~~$72\pm~4$&  2   &    6      &   1.64    &      0.0068      &$3.4\times 10^{-4}$\\
 4331 &Gas~~~$86\pm 5$&  1   &    4      &   1.37    &      0.0504      &                   \\
          &&&&&&\\
58547 &Gas~~~$99\pm~5$&  2   &    6      &   2.20    &      0.0248      &$7.9\times 10^{-5}$\\
23906 &Gas~~~$99\pm~5$&  2   &   10      &   3.48    &      0.0032      &                   \\
          &&&&&&\\
62192 &Gas~~$380\pm16$&  2   &    6      &   1.11    &     0.0010       &$6.6\times 10^{-8}$\\
20534 &Qua~~$930\pm28$&      &    6      &   2.37    &     0.0339       &                   \\
      &Gas~~$695\pm29$&  2   &    4      &   1.54    &     0.0708       &                   \\
      &Qua~$1024\pm31$&      &    5      &   1.67    &     0.0275       &                   \\
          &&&&&&\\
79940 &Gas~~$580\pm24$&  0   &    7      &   3.50    &     0.0661       &$7.2\times 10^{-7}$\\
 2890 &Liq~~$832\pm21$&      &    6      &   2.19    &     0.0246       &                   \\
      &Qua~~$990\pm30$&      &    5      &   1.72    &     0.0306       &                   \\
      &Gas~~$~83\pm~3$&  1   &    6      &   1.94    &     0.0145       &                   \\
          &&&&&&\\
%84432&Gas~~$~75\pm~2$&       &           &           &     0.0813       &$6.2\times 10^{-4}$\\
%53972&Gas~~$~75\pm~2$&       &           &           &     0.0076       &$                 $\\
%         &&&&&&\\
104949&Gas~~$442\pm17$&  2   &    8      &   4.59    &     0.0942       &$6.4\times 10^{-5}$\\
 27827&Liq~~$792\pm21$&      &    9      &   5.21    &     0.0827       &                   \\
      &Gas~~$742\pm31$&  0   &    6      &   3.07    &     0.0912       &                   \\
      &Liq~~$978\pm26$&      &    7      &   3.79    &     0.0900       &                   \\
          &&&&&&\\
\hline
\end{tabular}
\end{center}
\newpage
{\bf Table 5, continued.} Characteristics of anomalous rings found 
in events of topology~3.
\begin{center}
\begin{tabular}{ |c |c |c |c |c |c |c |}
\hline
          &&&&&&\\
 Run  &Ring radius,   &OD tag&No. of hits&  Expected & BG p-value      &Product of BG     \\
Event &   mrad        &      &  found    &  BG level &  per ring       &p-values per event\\
\hline
          &&&&&&\\

105892&Gas~~$104\pm~4$&  0   &    5      &   1.14    &     0.0063       &$1.8\times 10^{-5}$\\
 13127&Gas~~$104\pm~4$&  2   &    7      &   1.83    &     0.0029       &                   \\
         &&&&&&\\
114204&Gas~~$318\pm12$&  1   &   10      &   5.22    &     0.0407       &$4.1\times 10^{-7}$\\
 2668 &Liq~~$735\pm18$&      &   11      &   6.60    &     0.0726       &                   \\
      &Gas~~$875\pm35$&  1   &   11      &   4.53    &     0.0071       &                   \\
      &Liq~$1015\pm27$&      &    8      &   3.29    &     0.0195       &                   \\
          &&&&&&\\
\hline
\end{tabular}
\end{center}
\vskip0.6cm
\noindent Comments to Table 5. ``Qua" means the quartz radiator ring.
When it is preceded by a gas (liquid) ring this means that the quartz
ring pertain to the same track as the preceding gas (liquid) ring. 
 
Note, the quartz ring of a radius of 990~mrad, found in the event 79940:2890,
after recalculation of its radius to the expected liquid radiator ring radius
(823~mrad) and combined with the radius of the gaseous ring of 580~mrad
associated with the same track, gives additional entries (marked in dark blue)
to Figs.~\ref{fig:15}, \ref{fig:16}.

\newpage
\newpage
\begin{figure}
\vskip-2.5cm
\begin{center}
\epsfig{file=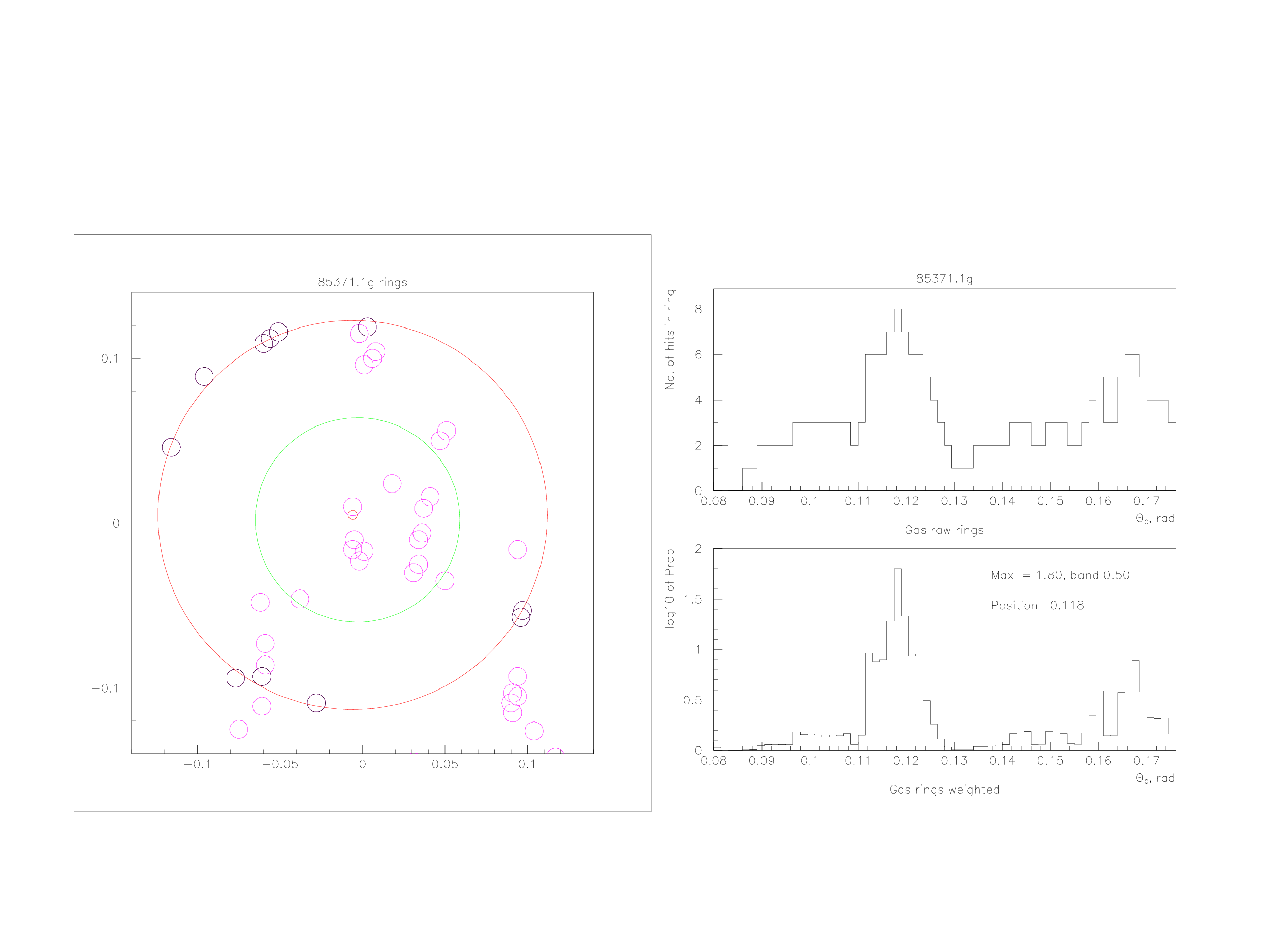,width=.84\textwidth}
\end{center}
\vskip-1.9cm
\caption{Gas radiator hit pattern for the 1st track of the event of topology 2
85371:626, a ring of a radius 118~mrad produced by the track, and the hit 
pattern radial distributions. Small circles in the hit pattern represent 
the RICH hits; their radii are equal to single photon angular accuracy 
$\sigma$ in the vicinity of the ring under consideration.
The hits pertaining to the anomalous ring are plotted in bold. 
The green circle marks the position and size of a standard ring 
(not seen in the pattern). The probability of anomalous ring (marked by the red
circle) to be fortuitously reconstructed from background hits is below 0.0158.
The units in the Cherenkov plane are given in radians, the small red circle in
its center marks the position of the image of the track impact point, 
$x=0,~y=0$.}
\label{fig:2}
\end{figure}

%\newpage
\begin{figure}
\vskip-0.4cm
\begin{center}
\epsfig{file=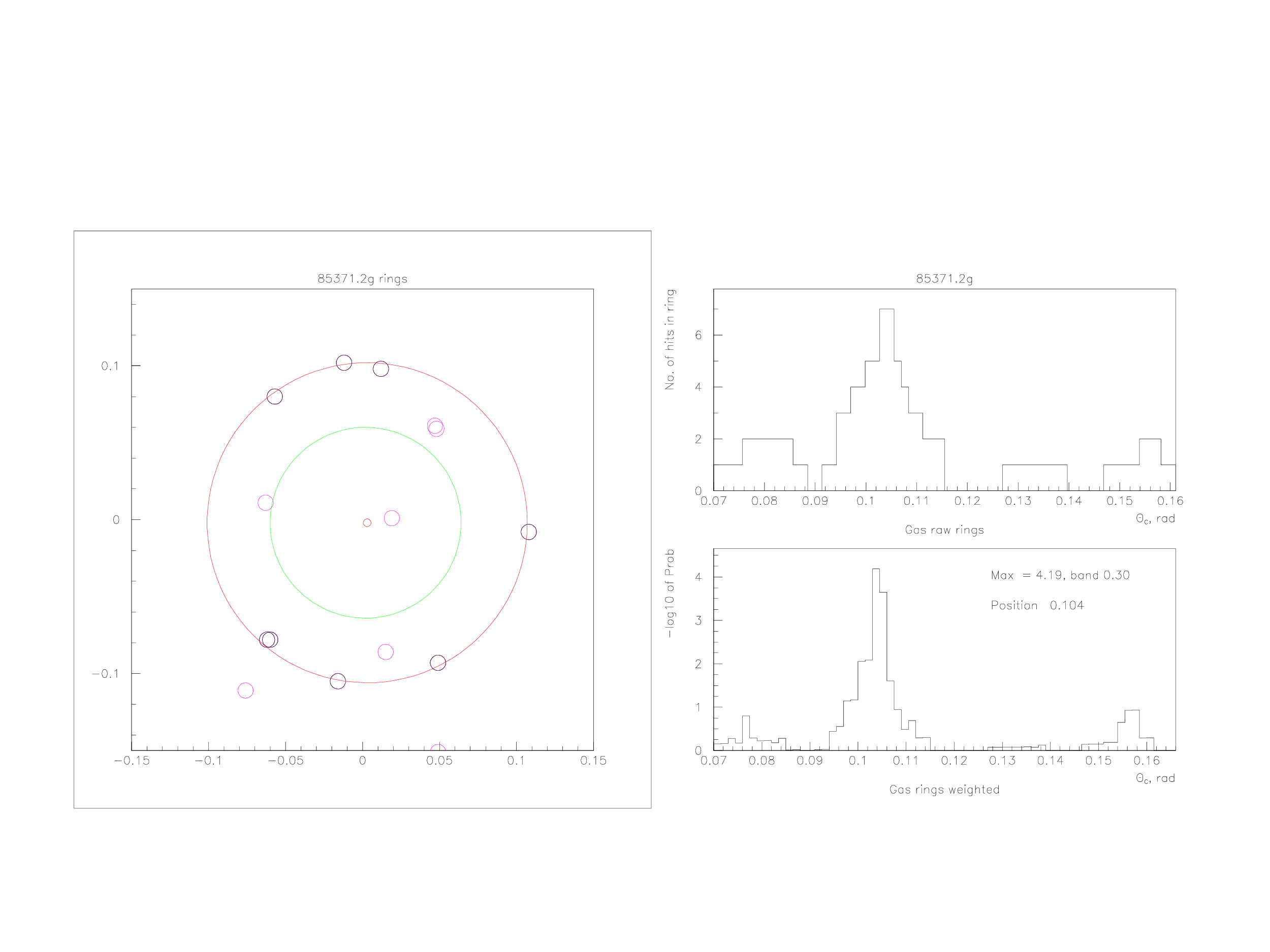,width=.84\textwidth}
\end{center}
\vskip-1.7cm
\caption{Gas radiator hit pattern for the 2nd track of the event 85371:626, 
a ring of a radius 104~mrad produced by the track, and the hit pattern radial 
distributions. The green circle marks the position and size of a standard ring 
(not seen in the pattern). The probability of anomalous ring (marked by the red
circle) to be fortuitously reconstructed from background hits is below 
$6.4\times10^{-5}$. The units in the Cherenkov plane are given in radians, 
the small red circle in its center marks the position of the image of the track 
impact point.}
\label{fig:3}
\end{figure}    

\begin{figure}
\begin{center}
\vskip-3.5cm
\epsfig{file=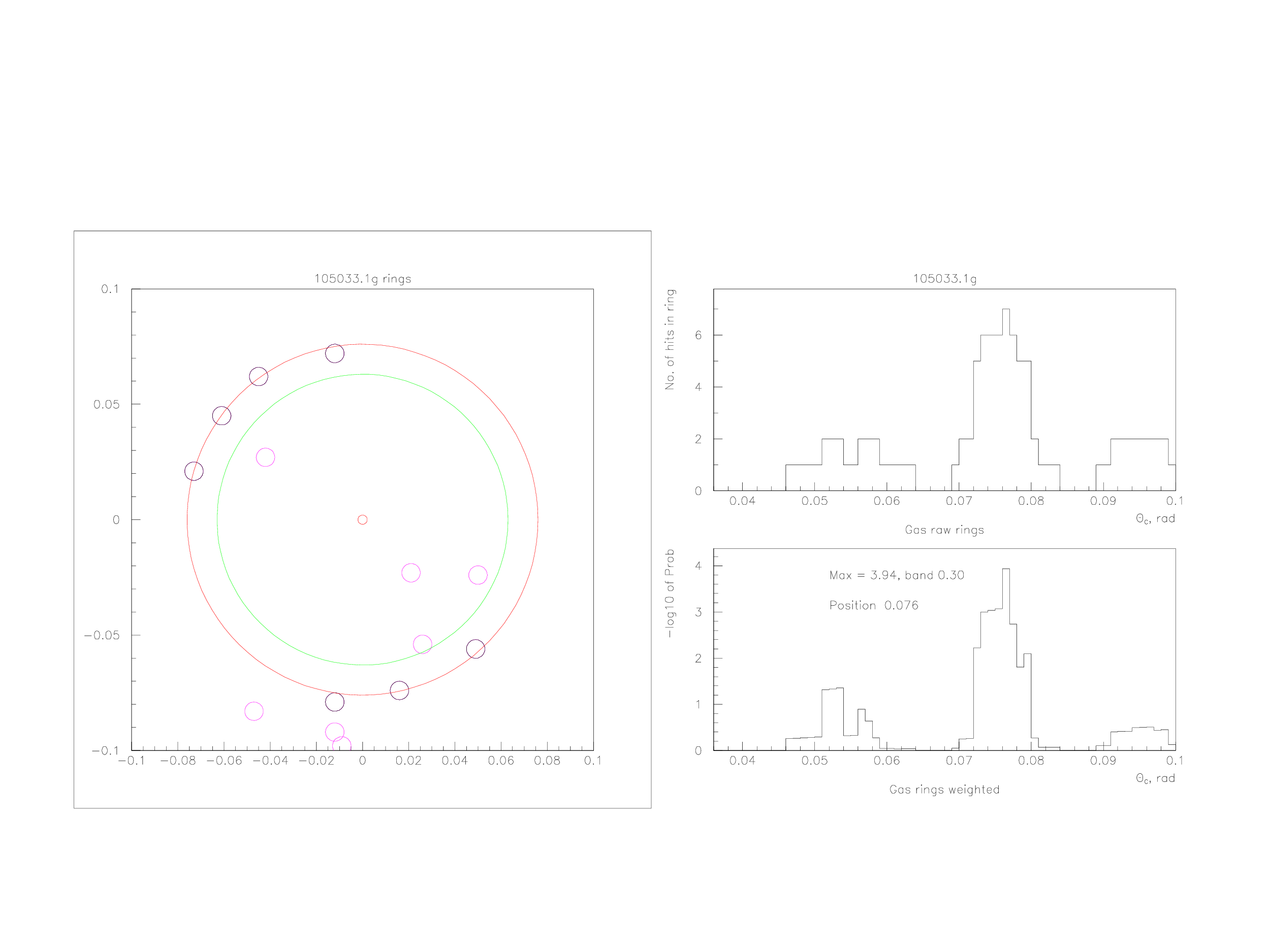,width=.82\textwidth}
\end{center}
\vskip-1.2cm
\caption{Gas radiator hit pattern for the 1st track of the event of topology 2
105033:3999, a ring of a radius~76 mrad produced by the track, and the hit 
pattern radial distributions. The green circle marks the position and size of 
a standard ring (not seen in the pattern). The probability of anomalous ring 
(marked by the red circle) to be fortuitously reconstructed from background 
hits is below $1.2\times10^{-4}$. The units in the Cherenkov plane are given in
radians, the small red circle in its center marks the position of the image
of the track impact point.}
\label{fig:4}
\end{figure}

\newpage
\begin{figure}
\begin{center}
\vskip-1.7cm
\epsfig{file=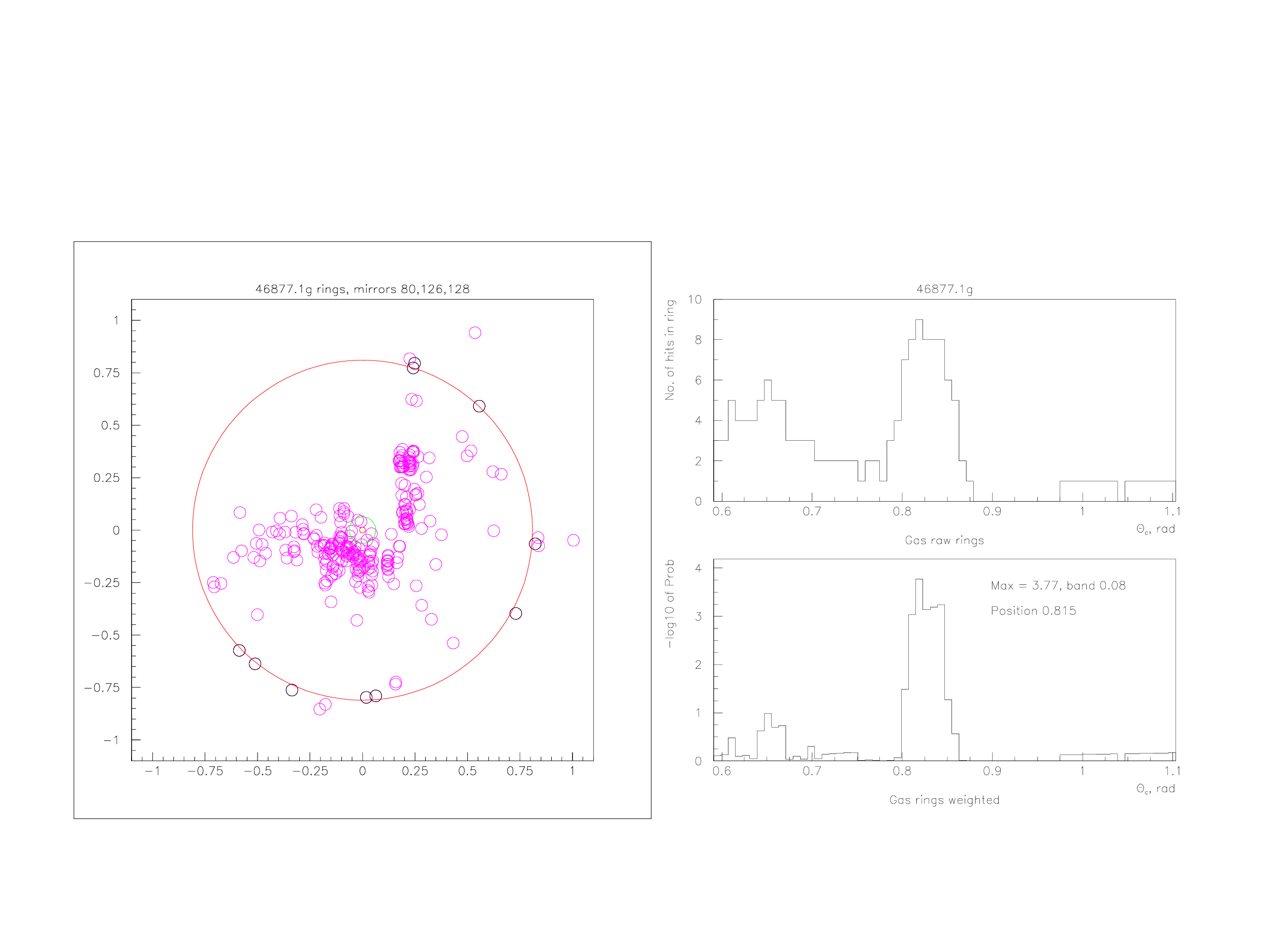,width=.82\textwidth}
\end{center}
\vskip-1.2cm
\caption{Gas radiator hit pattern for a track of the event of topology 3
46877:4278, a ring of a radius~815~mrad produced by the track, and the hit
pattern radial distributions. The green circle marks the position and size of
a standard ring (not seen in the pattern). The probability of anomalous ring
(marked by the red circle) to be fortuitously reconstructed from background
hits is below $1.7\times10^{-4}$. The units in the Cherenkov plane are given in
radians, the small red circle in its center marks the position of the image
of the track impact point.}
\label{fig:77}
\end{figure}

\begin{figure}
%\vskip-2cm
\begin{center}
\epsfig{file=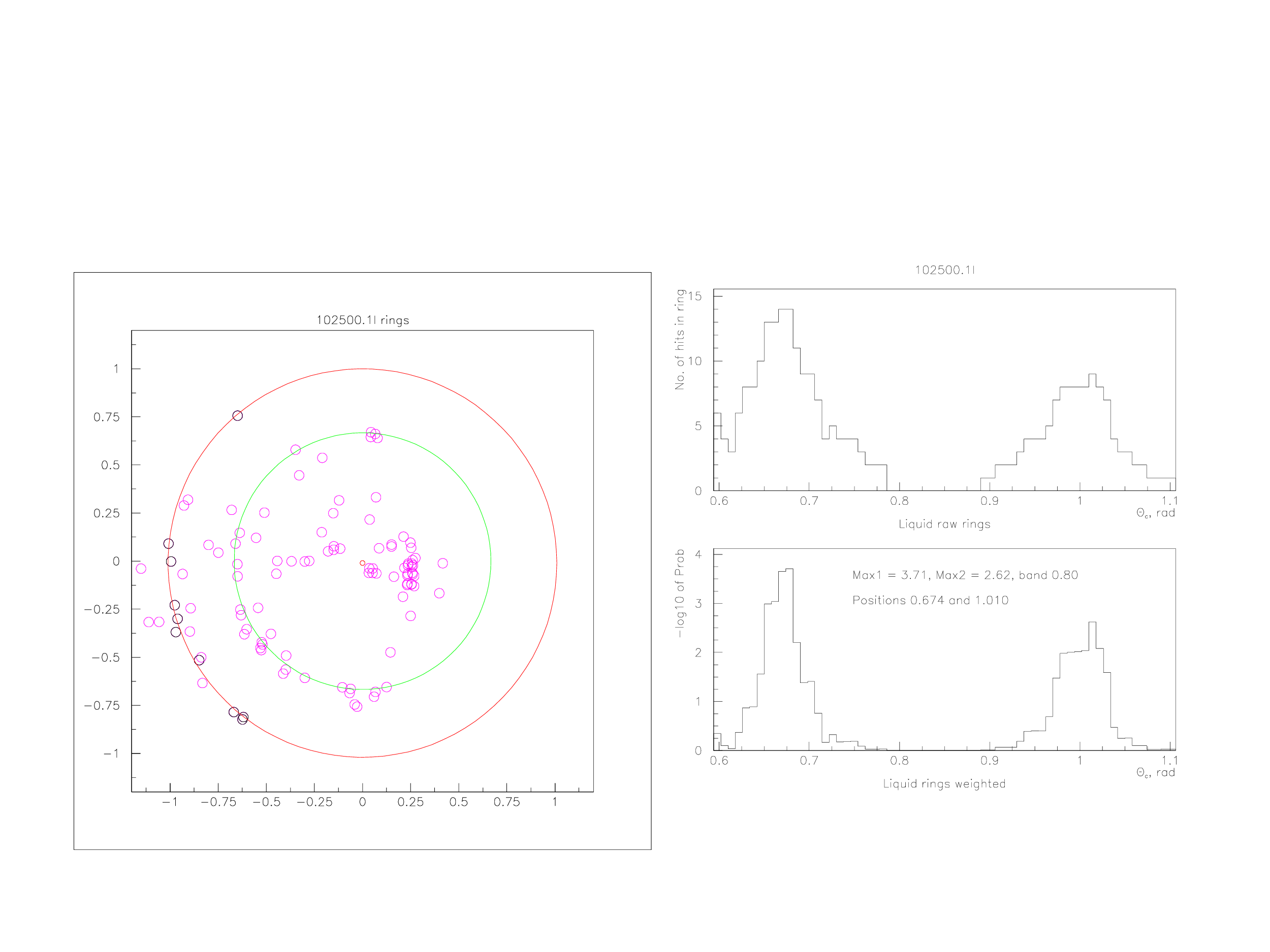,width=.82\textwidth}
\end{center}
\vskip-1.2cm
\caption{Liquid radiator hit pattern for a double track of an event of 
topology~1a 102500:9258. Together with an anomalous ring
(indeed, an arc due to the effect of the total inner reflection), having 
radius of 1010~mrad and marked by a red circle, a ring of a radius of 674~mrad
(close to the radius of a standard ring 667~mrad and also presented as an arc)
marked by a green circle, is seen. The probability of anomalous rings marked by
the red circle to be fortuitously reconstructed from background is below 
$2.4\times10^{-3}$. The units in the Cherenkov plane are given in radians, 
the small red circle in its center marks the position of the image of the 
track impact point.}
\label{fig:5}
\end{figure}
%\newpage
%\begin{figure}
%\begin{center}
%\epsfig{file=fig4.eps,bbllx=50pt,bblly=180pt,bburx=550pt,bbury=670pt,%
%width=17cm,angle=0}
%\end{center}
%\vskip1.5cm
%\caption{ The principle of the DELPHI Barrel RICH operation.}
%\label{fig:6}
%\end{figure}

\newpage
\begin{figure}
\begin{center}
\epsfig{file=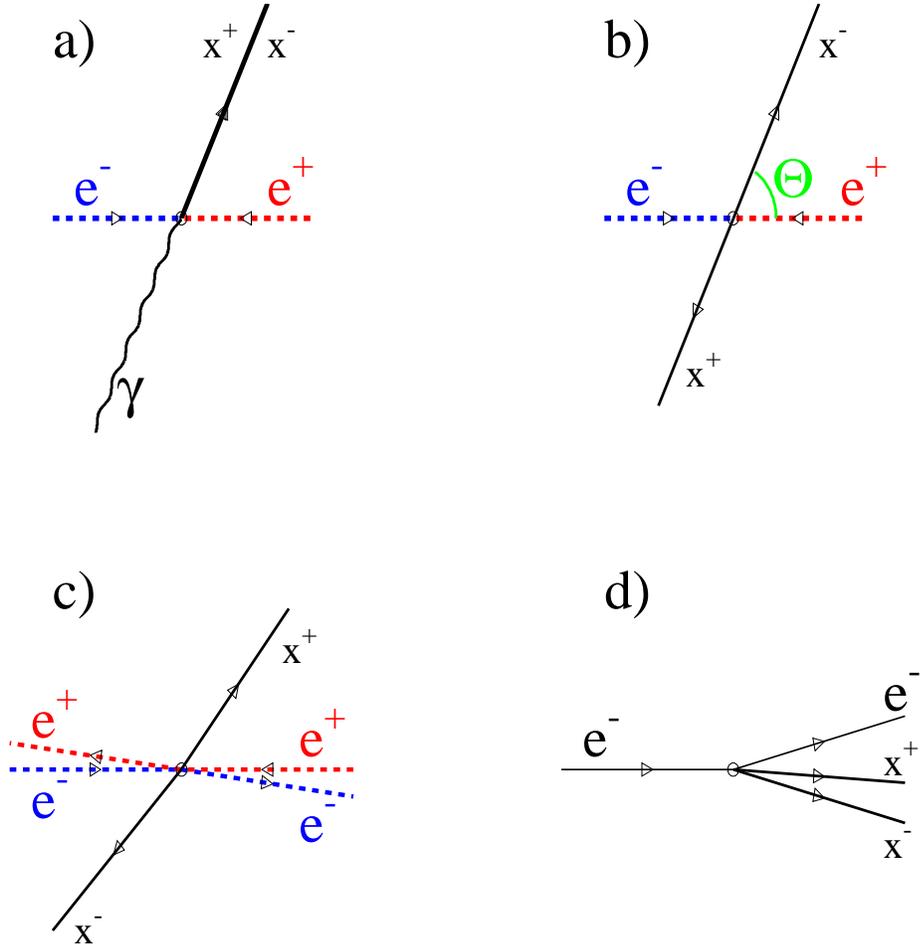,bbllx=0pt,bblly=200pt,bburx=650pt,bbury=570pt,%
width=19cm,angle=0}
\end{center}
%\vskip-1cm
\caption{Reaction diagrams of topologies studied: {\bf a)}~topology~1, 
production of an unlike-sign particle pair with nearly-zero opening angle; 
{\bf b)}~topology~2a, back-to-back particle production; 
{\bf c)}~topology~2b generated by a different mechanism of the particle pair 
production, via a $\gamma~\gamma$ interaction; 
{\bf d)}~topology~3, electroproduction of a particle pair on nuclei in the
detector material. The dashed lines mark beam particles and/or the particles 
that go undetected escaping into the beam pipe.}
\label{fig:6}
\end{figure}    

\newpage
\begin{figure}
\begin{center}
\epsfig{file=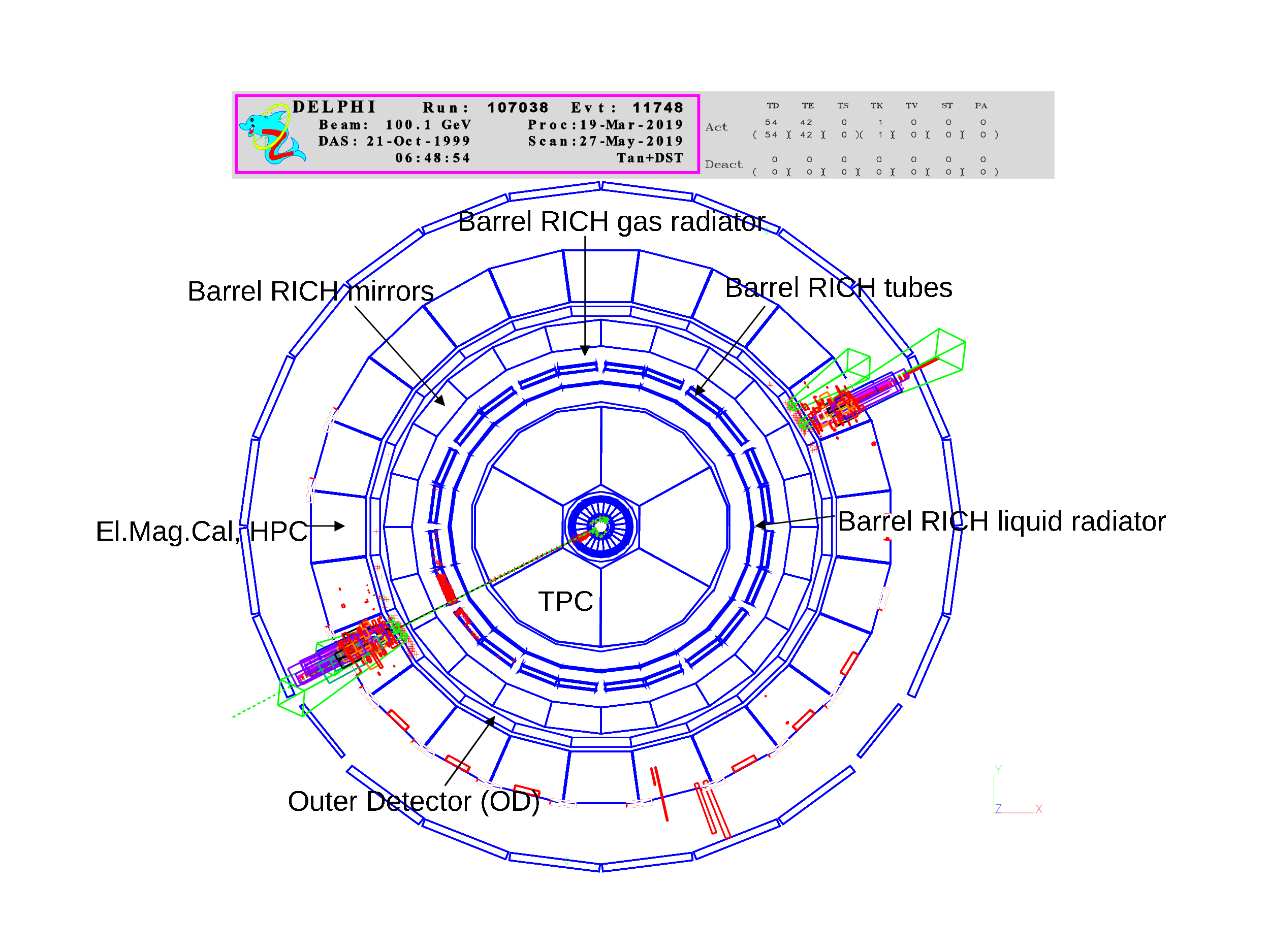,bbllx=100pt,bblly=000pt,bburx=675pt,bbury=400pt,%
width=17cm,angle=0}
\end{center}
\vskip 1cm
\caption{ An example of event of topology 1a of reaction (3.1) with two tracks 
non-resolved in the TPC, a general view. 
The red hits inside the Barrel RICH drift tubes are Barrel RICH hits (the
tubes will not be shown in next figures).}
\label{fig:61}
\end{figure}

\newpage
\begin{figure}
\begin{center}
\epsfig{file=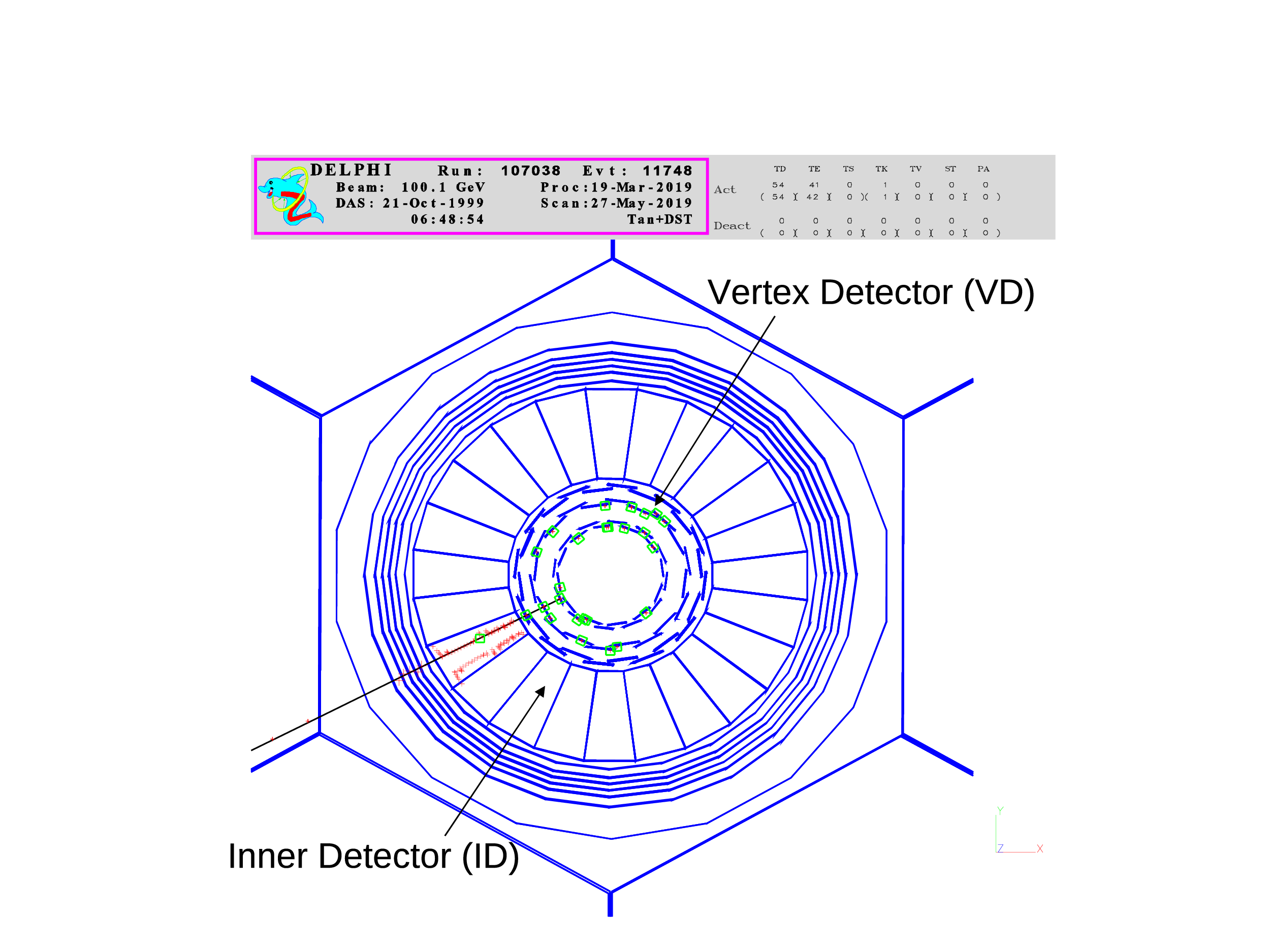,bbllx=100pt,bblly=00pt,bburx=675pt,bbury=500pt,%
width=19cm,angle=0}
\end{center}
\vskip 1cm
\caption{ The same event, the central detector (VD and ID) view.}
\label{fig:161}
\end{figure}

\newpage
\begin{figure}
\begin{center}
\epsfig{file=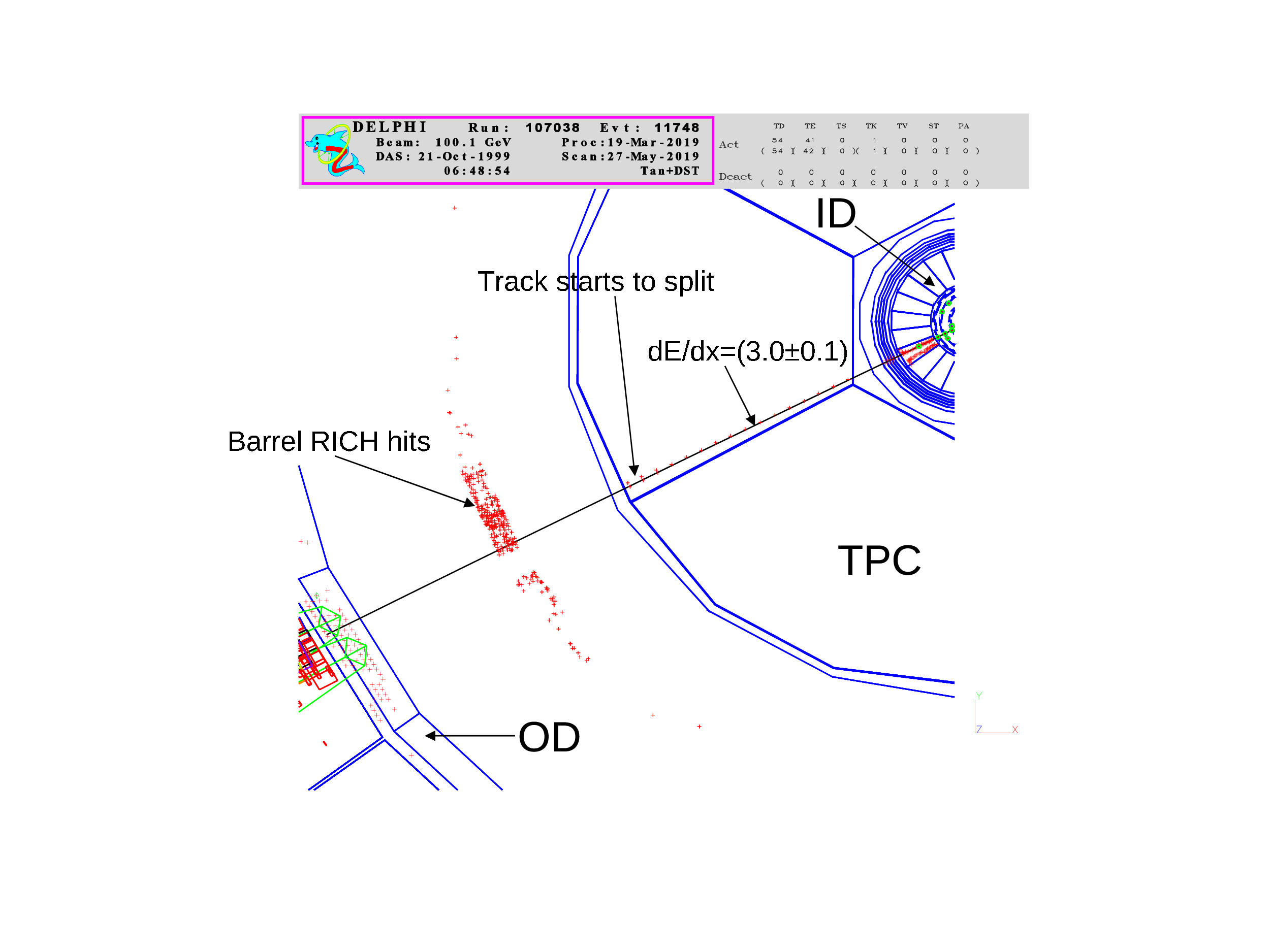,bbllx=100pt,bblly=0pt,bburx=675pt,bbury=500pt,%
width=19cm,angle=0}
\end{center}
%\vskip 3cm
\caption{The same event, the TPC view showing the non-resolved track starting 
to split. The $dE/dx$ units are mips (see footnote 3 on page~4).} 
\label{fig:261}
\end{figure}

\newpage
\begin{figure}
\begin{center}
\epsfig{file=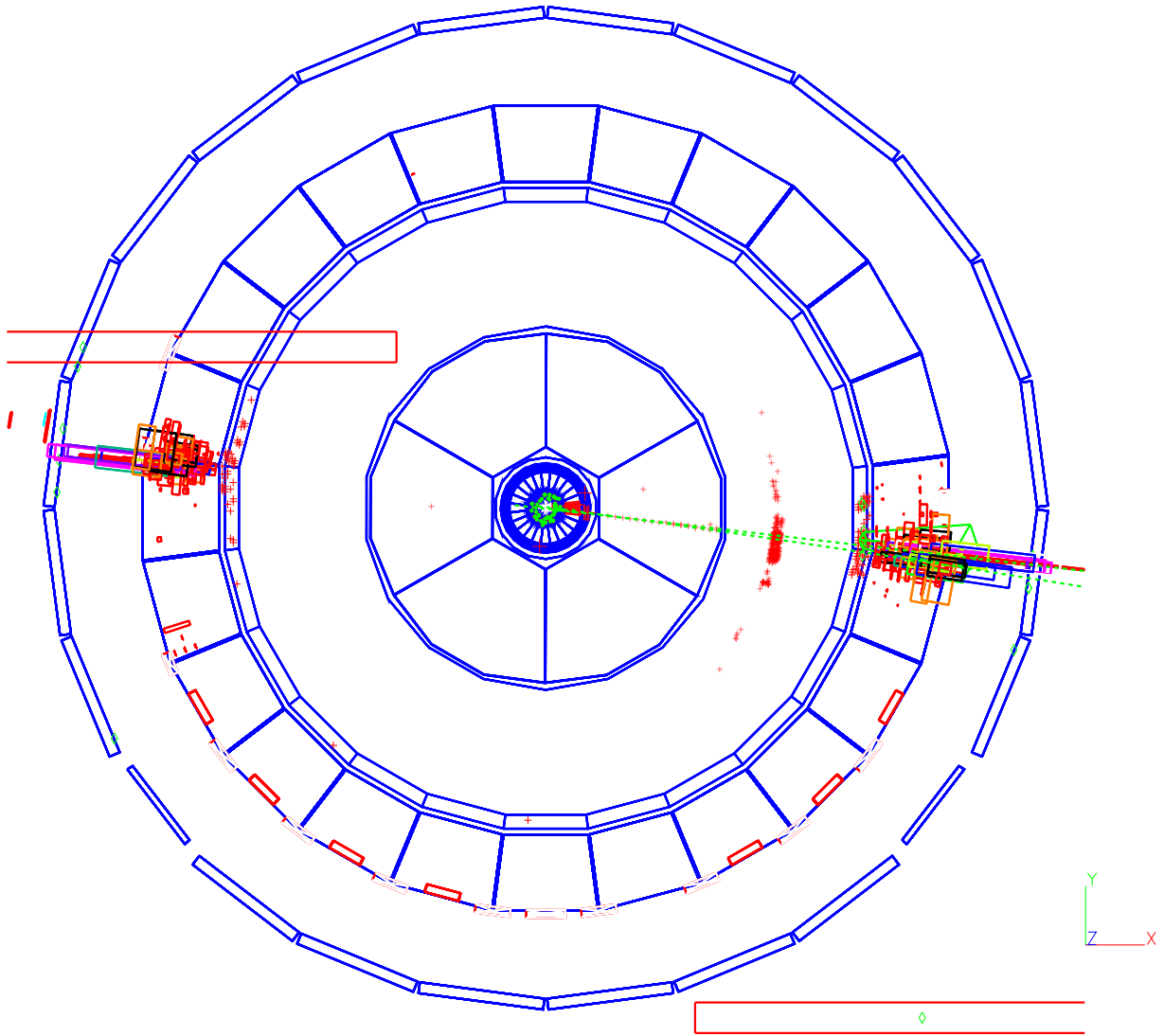,bbllx=0pt,bblly=0pt,bburx=675pt,bbury=700pt,%
width=20cm,angle=0}
\end{center}
\vskip-2cm
\caption{ An example of event of topology 1b of reaction (3.1) with two tracks 
resolved in the TPC, a general view.}
\label{fig:61a}
\end{figure}

\newpage
\begin{figure}
\begin{center}
\epsfig{file=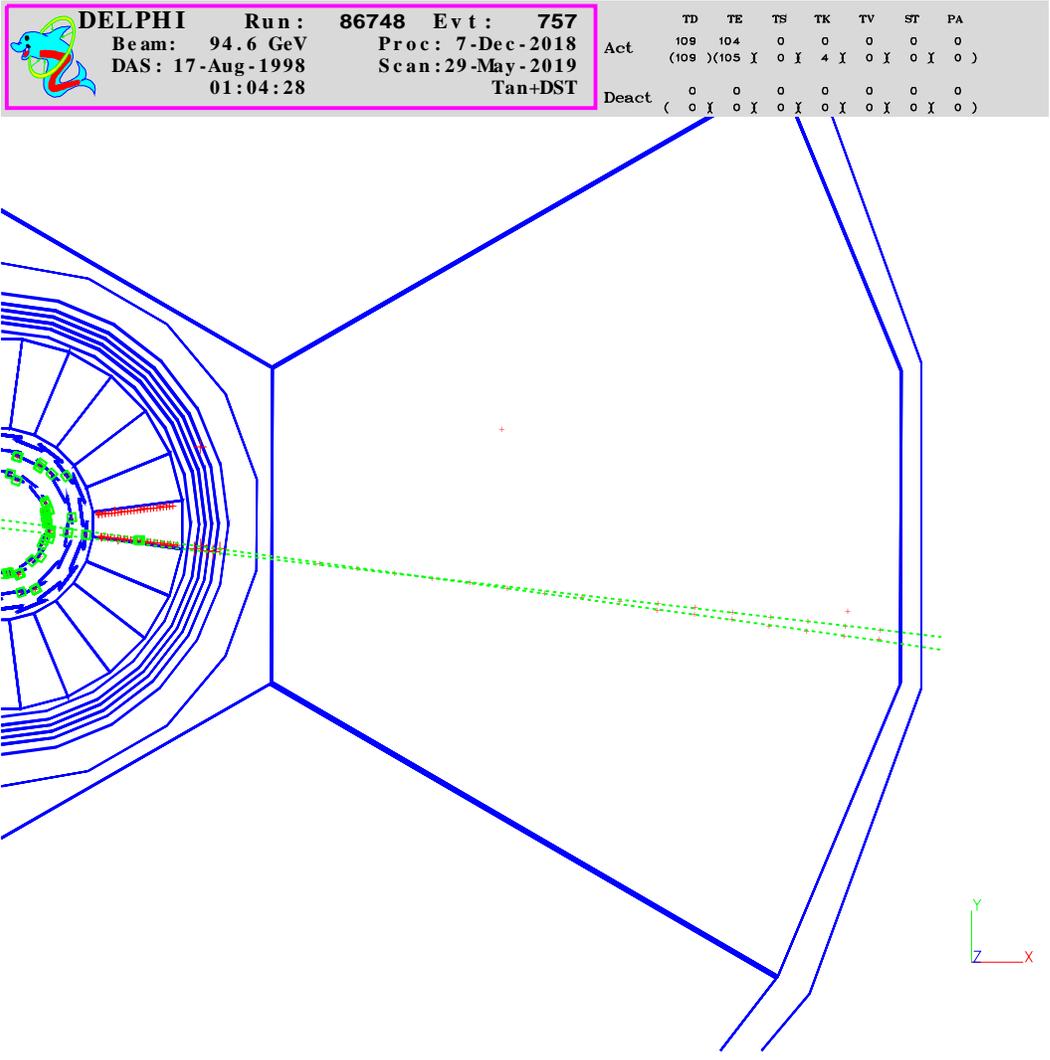,bbllx=0pt,bblly=0pt,bburx=675pt,bbury=500pt,%
width=17cm,angle=0}
\end{center}
\vskip-2cm
\caption{ The same event, the VD, ID and TPC view of the two-particle resolved 
jet.}
\label{fig:161a}
\end{figure}

% go to topology 2
\newpage
\begin{figure}
\begin{center}
\epsfig{file=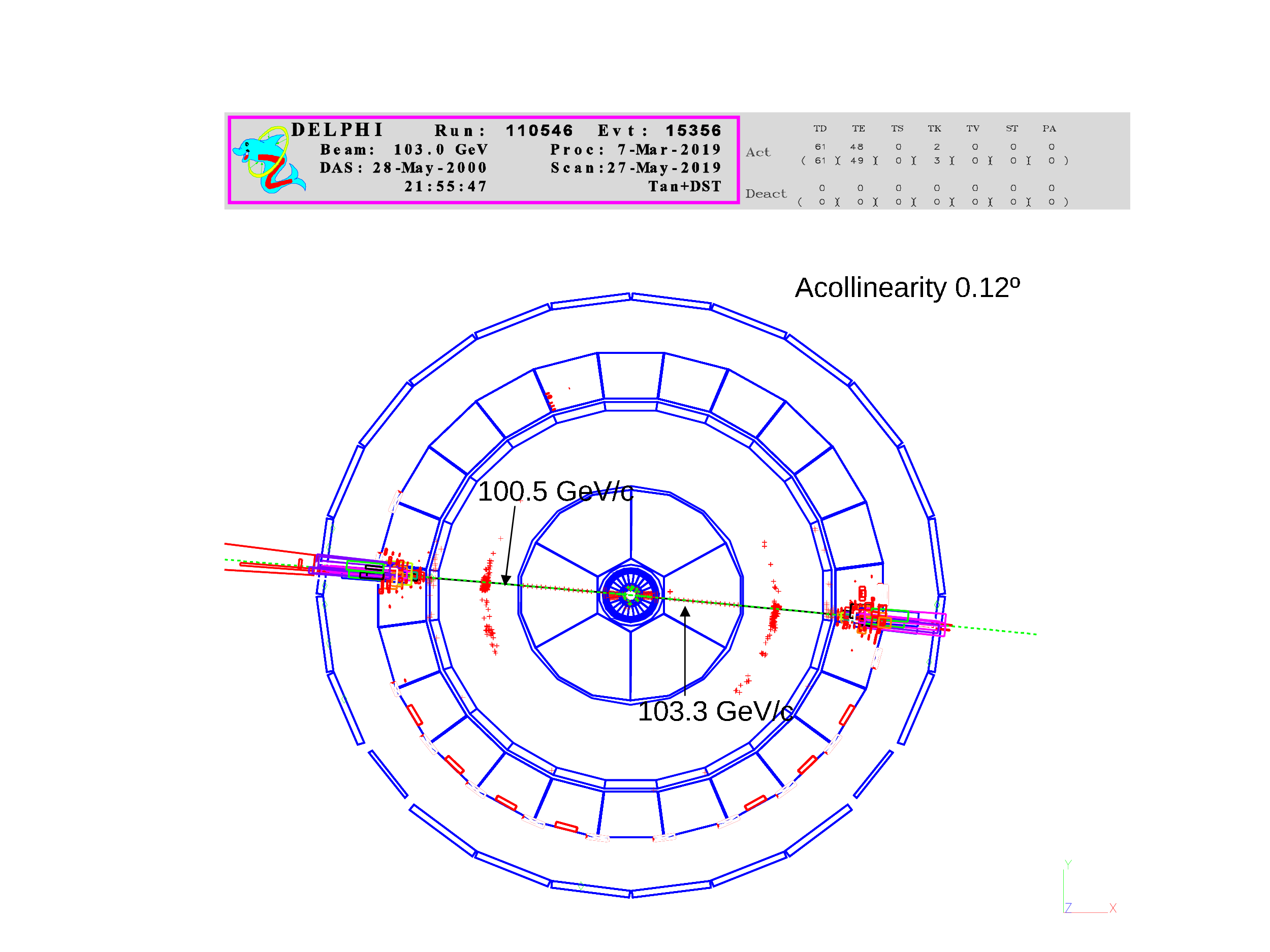,bbllx=120pt,bblly=0pt,bburx=675pt,bbury=500pt,%
width=18cm,angle=0}
\end{center}
\vskip3cm
\caption{ An example of event of topology 2a of reaction 3.2, a general view.}
\label{fig:62a}
\end{figure}

\newpage
\begin{figure}
\begin{center}
\epsfig{file=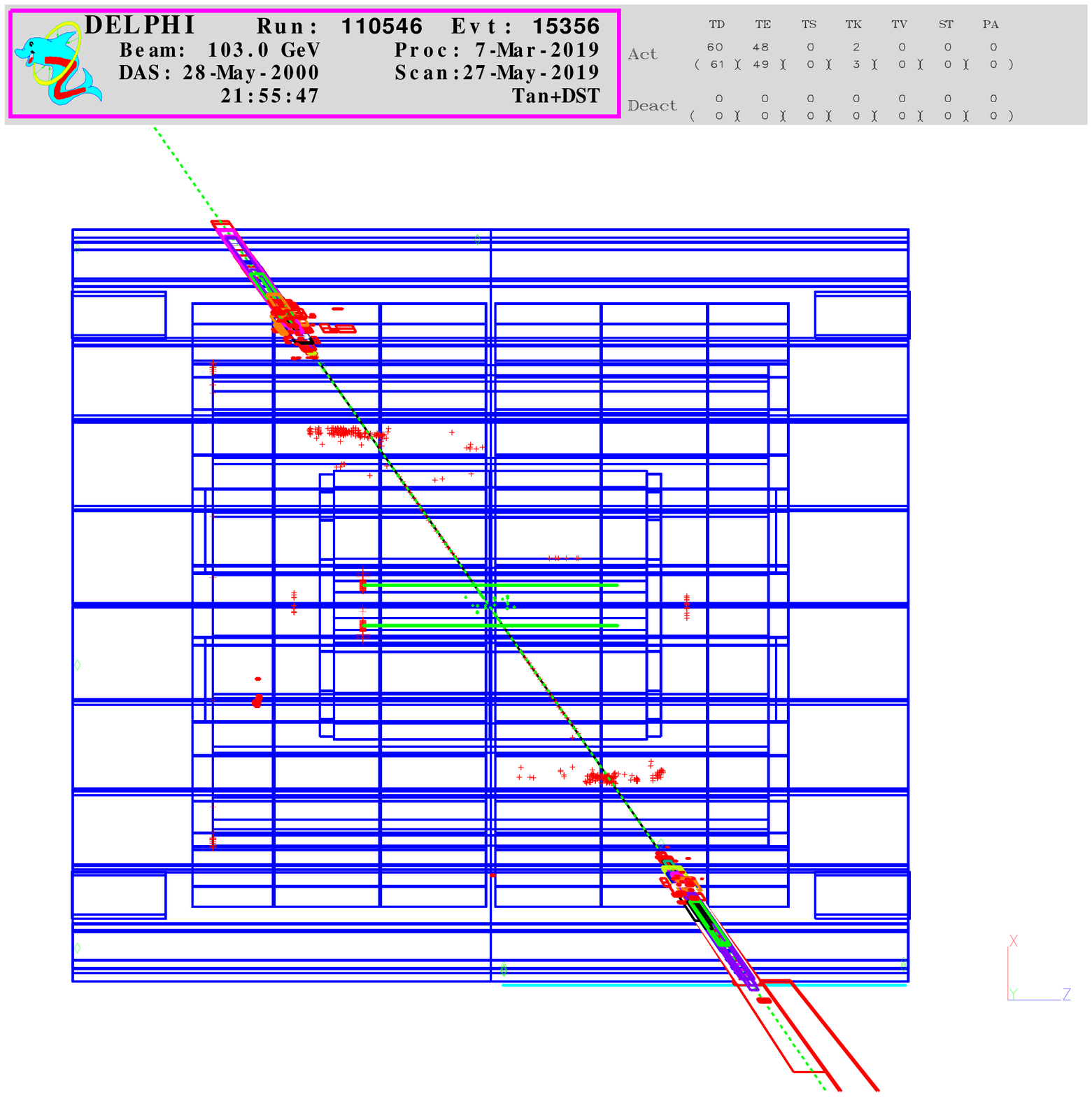,bbllx=0pt,bblly=0pt,bburx=675pt,bbury=800pt,%
width=20cm,angle=0}
\end{center}
\vskip-3cm
\caption{ The same event, the XZ view.}
\label{fig:162a}
\end{figure}

\newpage
\clearpage
\begin{figure}
\begin{center}
\epsfig{file=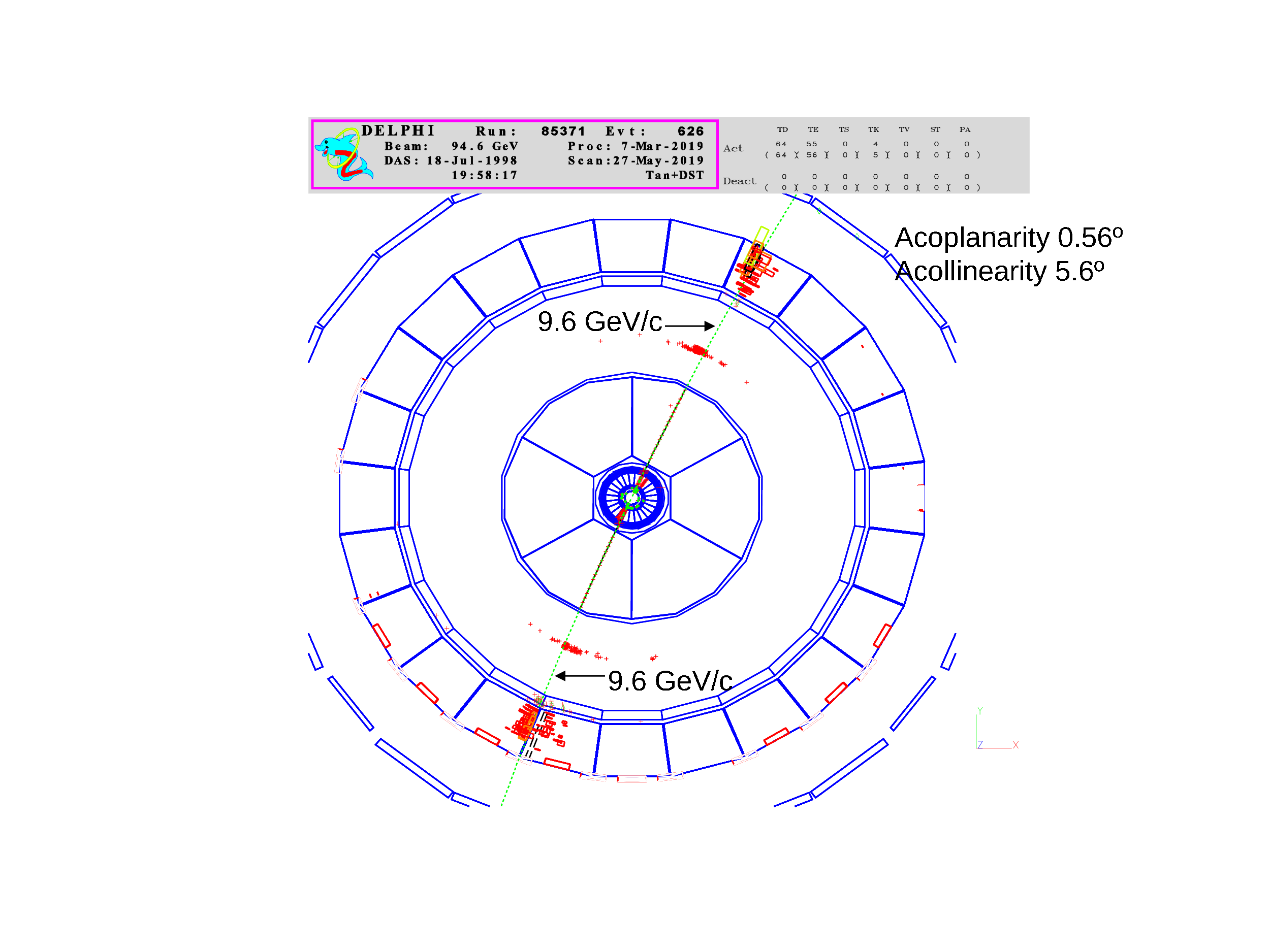,bbllx=120pt,bblly=0pt,bburx=675pt,bbury=500pt,%
width=18cm,angle=0}
\end{center}
\vskip 1cm
\caption{ An example of event of topology 2b of reaction 3.3, a general view.}
\label{fig:62}
\end{figure}

\newpage
\clearpage
\begin{figure}
\begin{center}
\epsfig{file=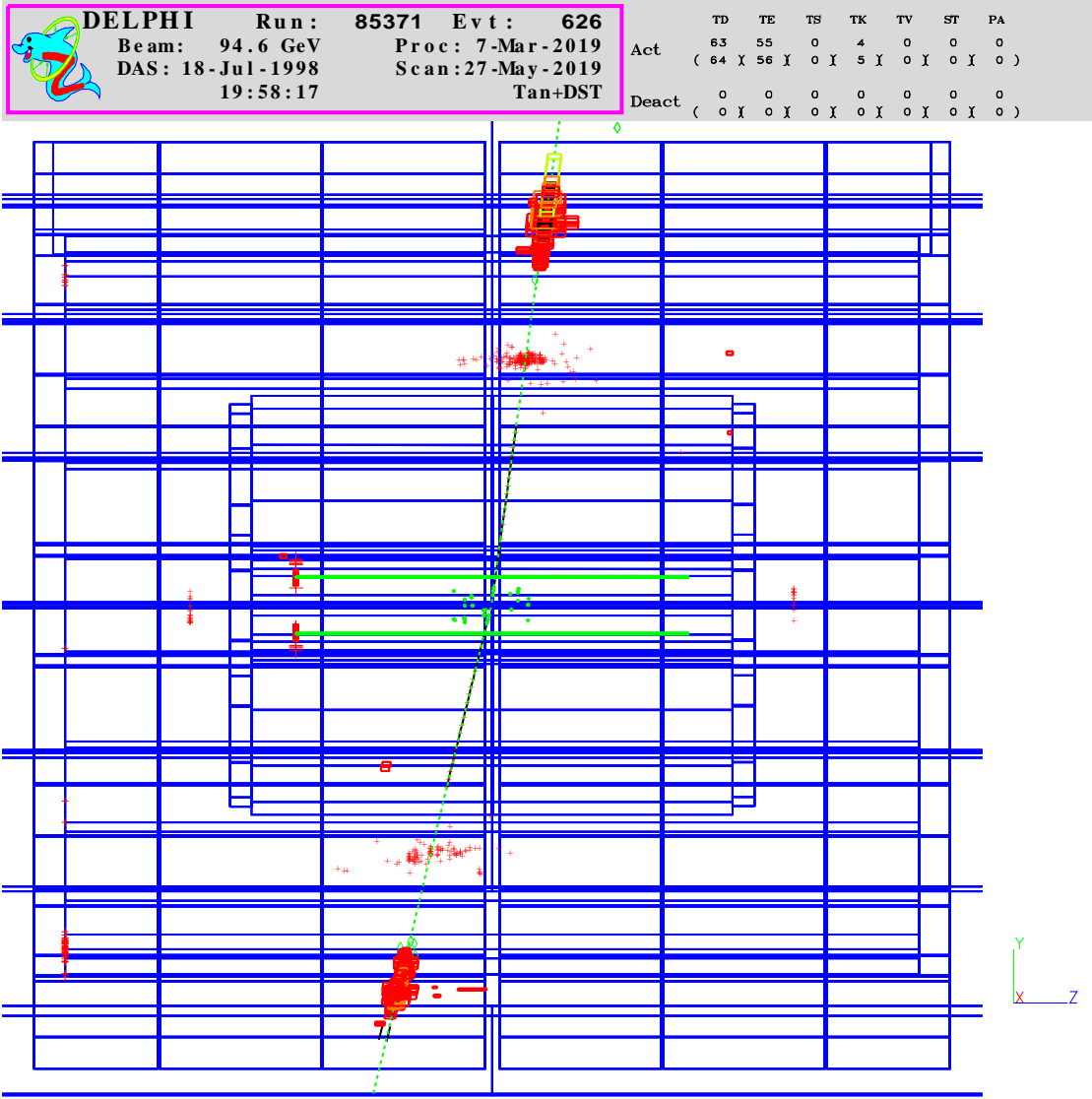,bbllx=0pt,bblly=0pt,bburx=675pt,bbury=700pt,%
width=18cm,angle=0}
\end{center}
\vskip 1cm
\caption{ The same event, the YZ view.}
\label{fig:162}
\end{figure}

%\newpage
%\begin{figure}
%\begin{center}
%\epsfig{file=85371gen.ps,bbllx=150pt,bblly=0pt,bburx=675pt,bbury=570pt,%
%width=20cm,angle=0}
%\end{center}
%\vskip-3cm
%\caption{ Another example of event of topology 2b, a general view.}
%\label{fig:62a}
%\end{figure}

\newpage
\begin{figure}
\begin{center}
\epsfig{file=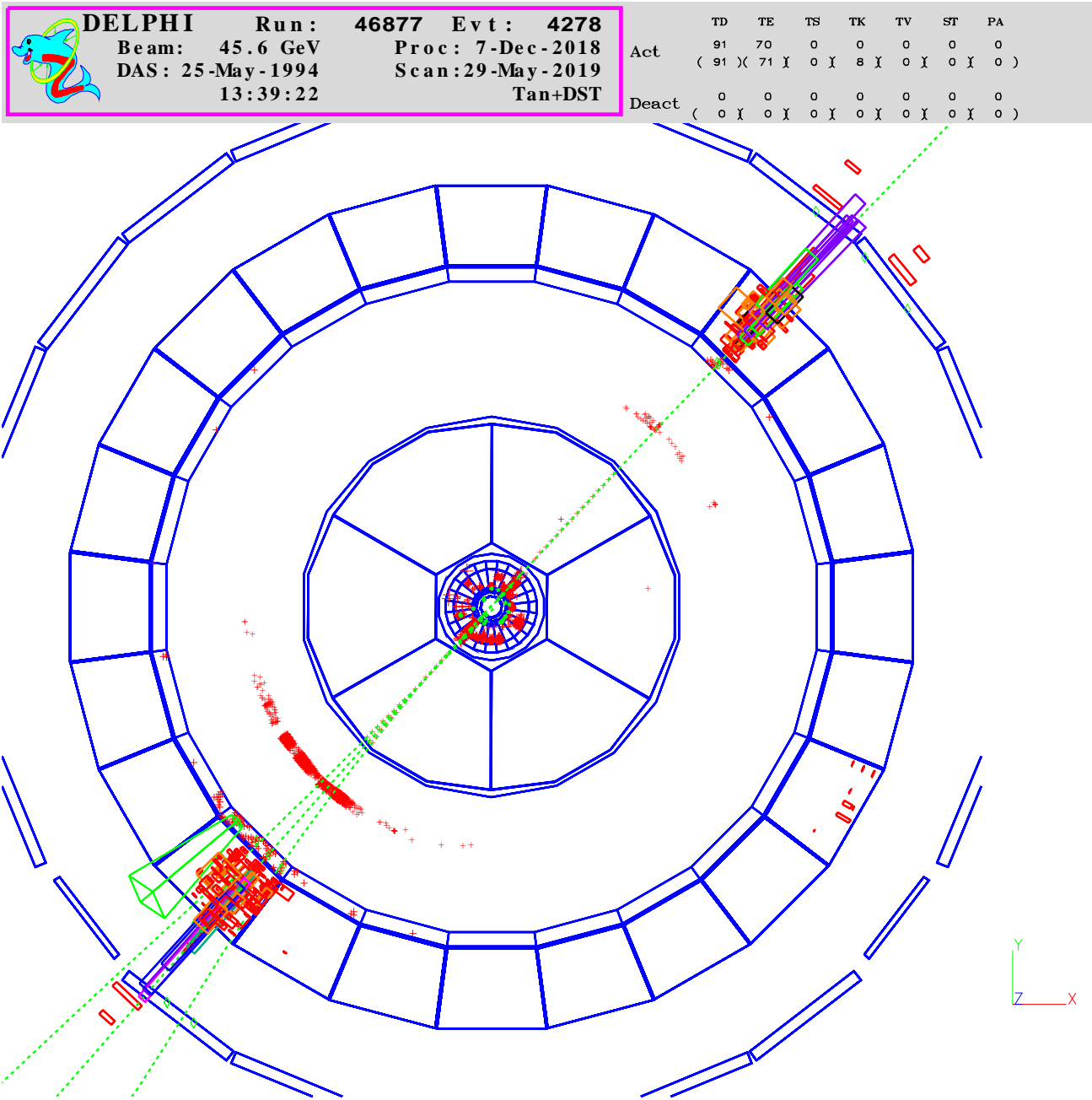,bbllx=50pt,bblly=250pt,bburx=500pt,bbury=600pt,%
width=14cm,angle=0}
\end{center}
\vskip5cm
\caption{ An event of topology 3, a general view.}
\label{fig:63}
\end{figure}

\newpage
\begin{figure}
\begin{center}
\epsfig{file=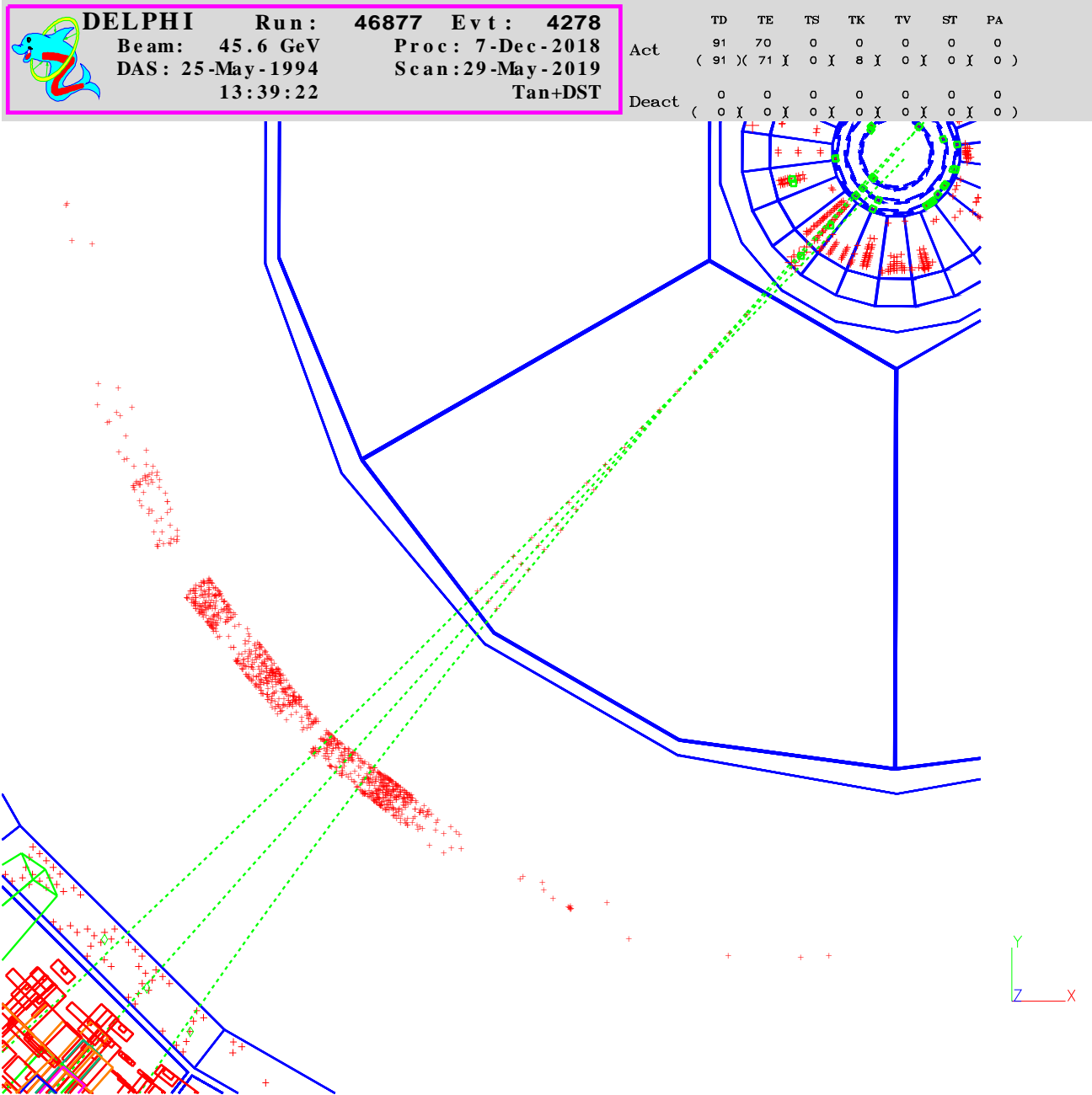,bbllx=50pt,bblly=250pt,bburx=500pt,bbury=600pt,%
width=14cm,angle=0}
\end{center}
\vskip5cm
\caption{ The same event, the 3-particle jet view.}
\label{fig:163}
\end{figure}

%\newpage
%\begin{figure}
%\begin{center}
%\epsfig{file=1_r105892_e013127.eps,bbllx=50pt,bblly=250pt,bburx=500pt,bbury=600pt,%
%width=14cm,angle=0}
%\end{center}
%\vskip5cm
%\caption{ Another event of topology 3, a general view.}
%\label{fig:63a}
%\end{figure}
 
\newpage
\begin{figure}
\begin{center}
\epsfig{file=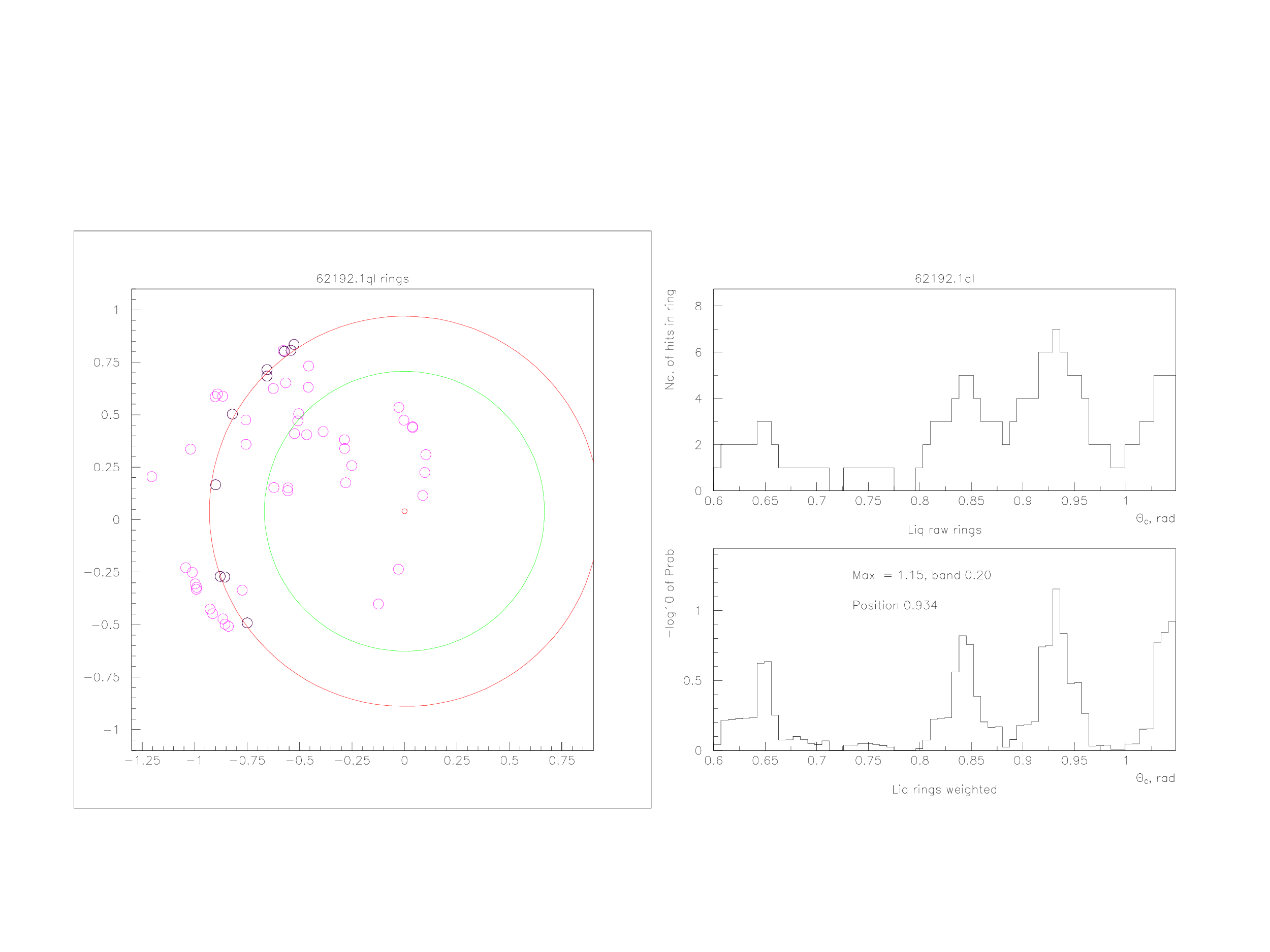,width=.80\textwidth}
\end{center}
\vskip-1cm
\caption{Quartz radiator hit pattern, obtained with the 1st track of the
event 62192:20534 of topology~3, an arc produced by it, and its radial 
distributions. The green circle shows the position of the standard liquid ring 
(not seen in the pattern).}
\label{fig:13}
\end{figure} 

\newpage
\begin{figure}
\begin{center}
\epsfig{file=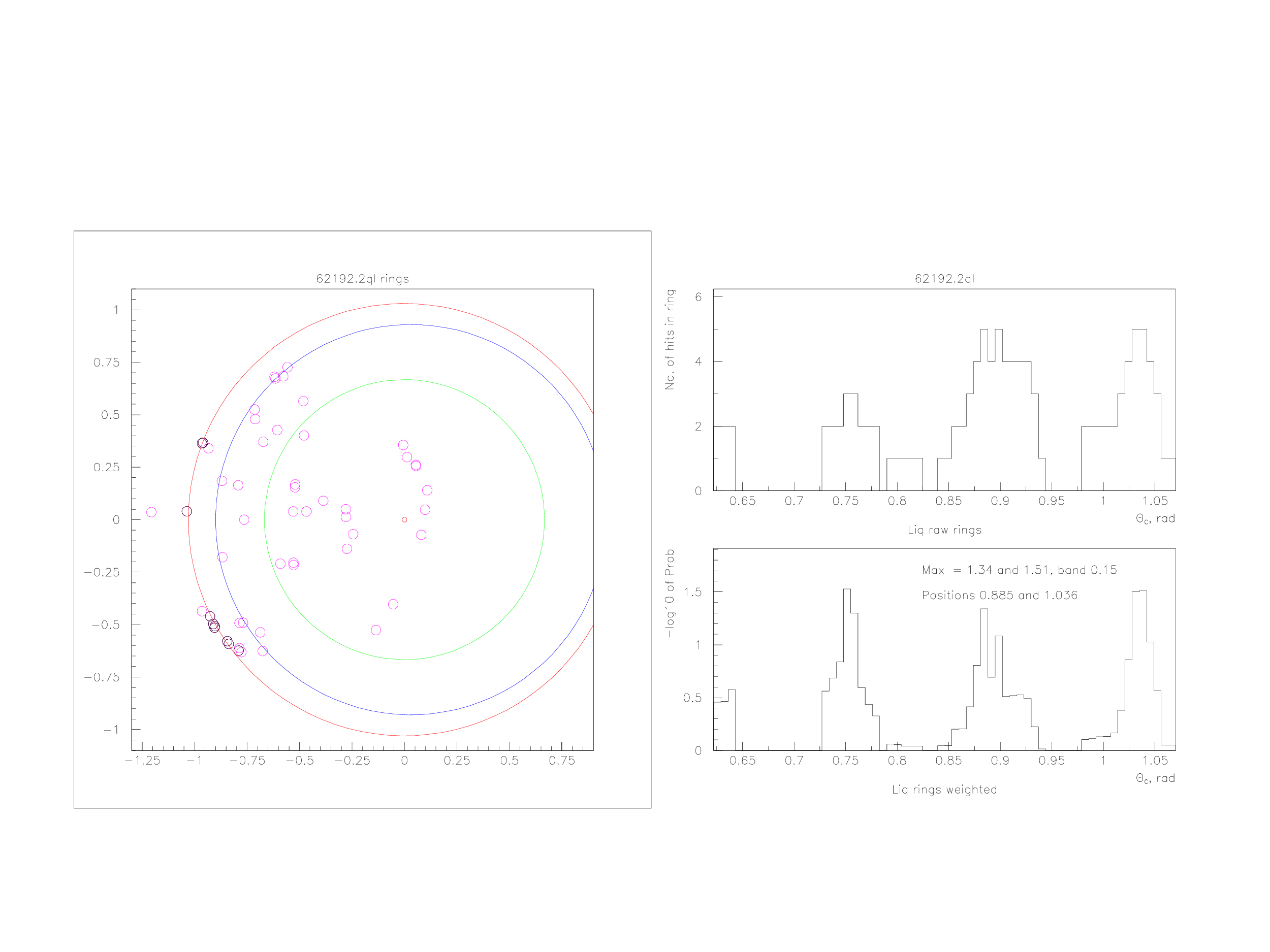,width=.80\textwidth}
\end{center}
\vskip-1.5cm
\caption{Quartz radiator hit pattern, obtained with the 2nd track of the
same event 62192:20534, an arc produced by it, and its radial distributions. 
The hits on the circle plotted in dark blue belong to the anomalous ring 
associated with the 1st track of the event, to be compared with the hits on 
the red circle in Fig.~\ref{fig:13}. The difference between the hit patterns
shown in Fig.~\ref{fig:13} and Fig.~\ref{fig:14} is due to the difference 
of the corresponding track directions.}
\label{fig:14}
\end{figure}

\newpage
\begin{figure}
\begin{center}
\epsfig{file=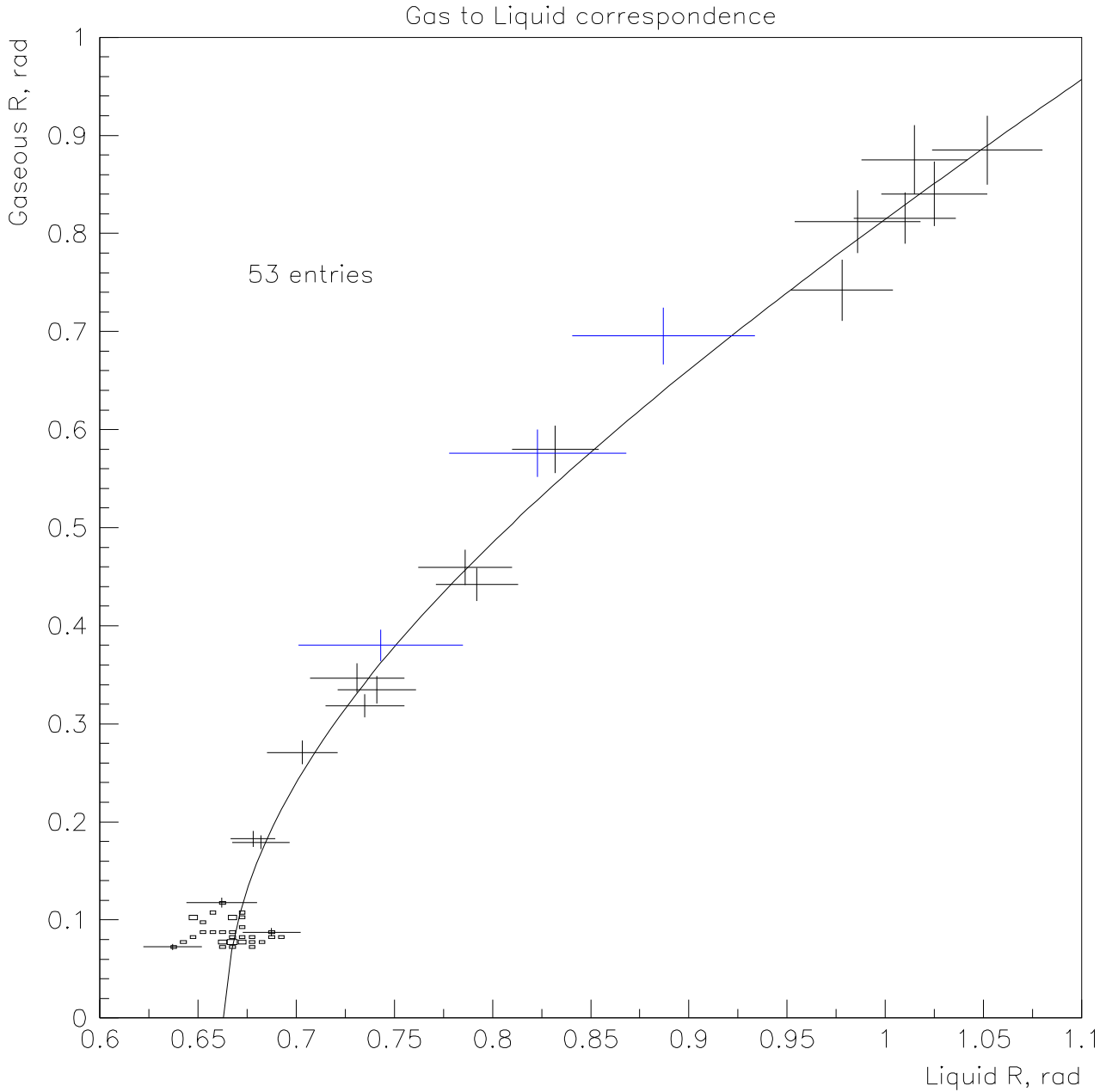,bbllx=50pt,bblly=180pt,bburx=550pt,bbury=670pt,%
width=17cm,angle=0}
\end{center}
\vskip0.3cm
\caption{ Radii of the anomalous rings expressed in angular units
as measured in the liquid and in the gaseous radiators. The curved line 
is not a fit to the data but shows the expected theoretical correspondence 
between the radii in the two radiators
given by the formula $n_{liq} \cos{\theta_{liq}} = n_{gas} \cos{\theta_{gas}}$,
with $n_{liq} = 1.273$ and $n_{gas} = 1.00194$. The errors in the left bottom
corner are suppressed to avoid pile-up; only several crosses are left in order
to present a typical angular error in this region.
3 crosses drawn in dark blue correspond to anomalous rings coming from the
quartz radiator. Their radii are recalculated to the liquid ring radii taking
into account the difference of the corresponding refraction indices.
For presentation purposes the scale along the $x$ axis is made two times finer 
than that of the $y$ axis.}
\label{fig:15}
\end{figure}

\newpage
\begin{figure}
\begin{center}
\epsfig{file=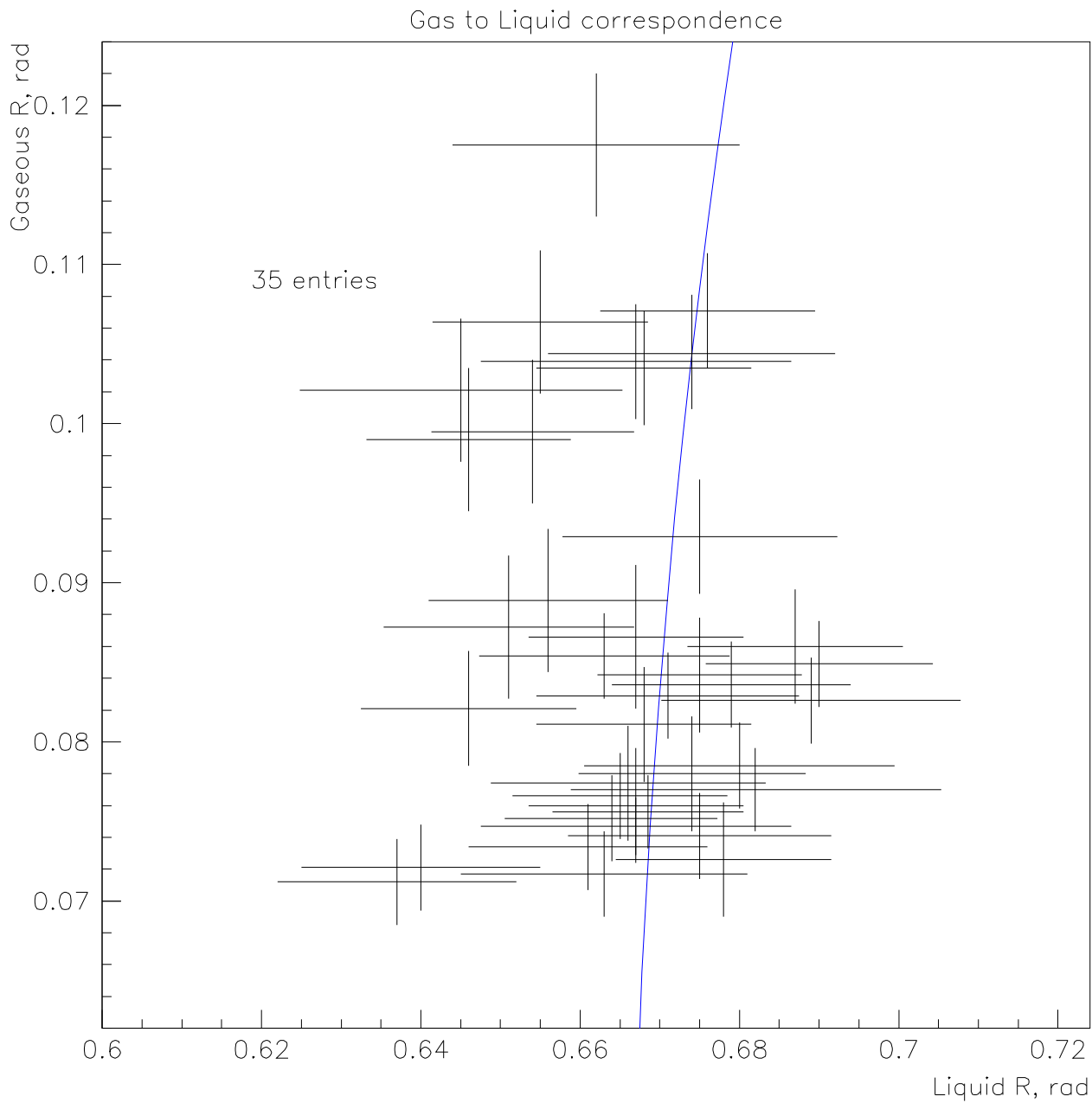,bbllx=50pt,bblly=180pt,bburx=550pt,bbury=670pt,%
width=17cm,angle=0}
\end{center}
\vskip1cm
\caption{ Radii of the anomalous rings as measured in the liquid and in the
gaseous radiators from the left bottom corner of Fig.~\ref{fig:15}. 
The expected theoretical correspondence between the radii
in the both radiators is shown by a curve. 
For presentation purposes the scale along the $y$ axis is made two times finer 
than that of the $x$ axis.}
\label{fig:15a}
\end{figure}

\newpage
\begin{figure}
\begin{center}
\epsfig{file=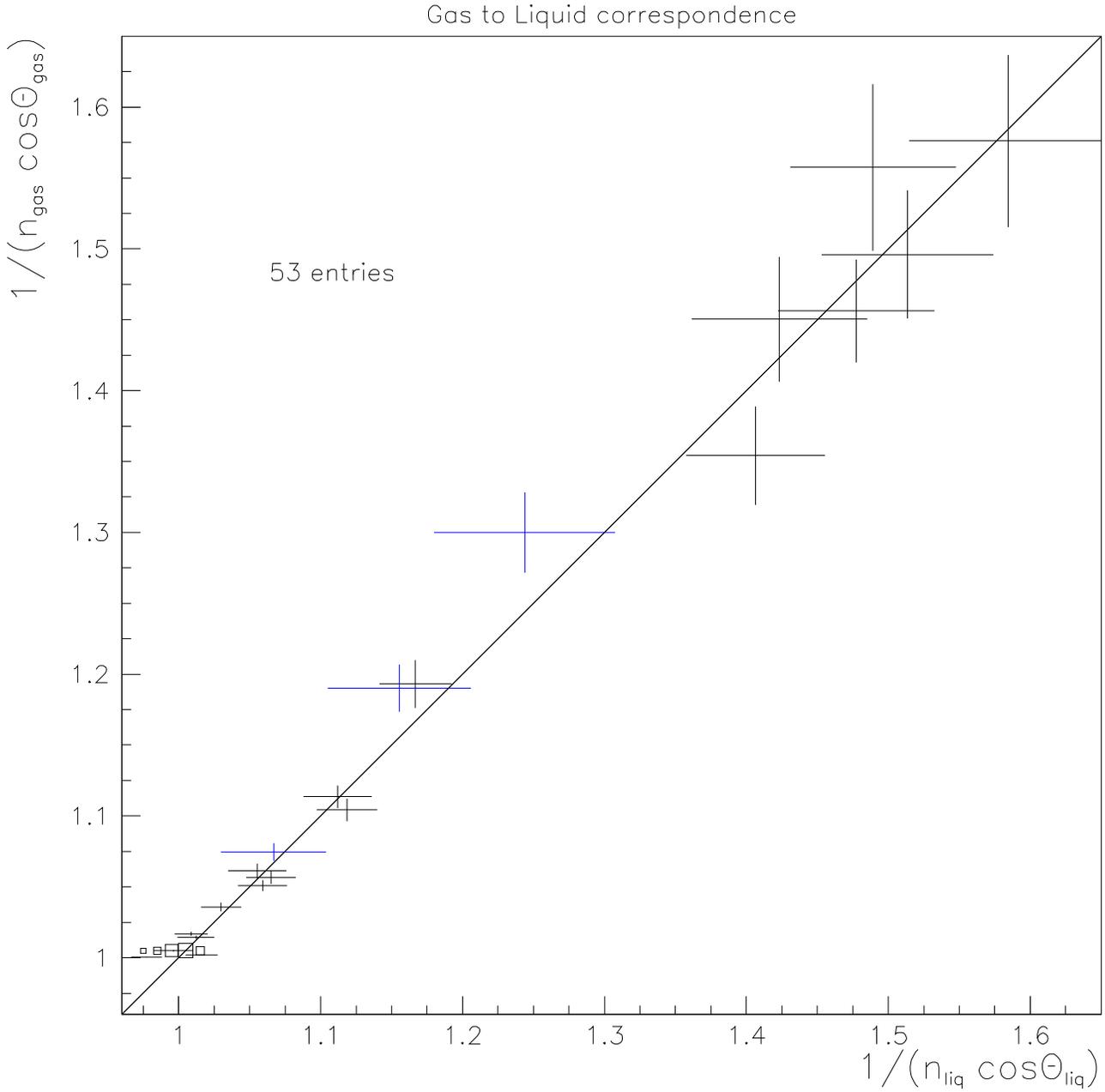,bbllx=50pt,bblly=180pt,bburx=550pt,bbury=670pt,%
width=17cm,angle=0}
\end{center}
\vskip2cm
\caption{Linearized correlation plot. 
%Particle velocities ($\beta = v/c$) corresponding to radii of 
%the anomalous rings as measured in the liquid and in the gaseous radiators. 
The diagonal indicates where the values of the two variables 
$1/(n_{liq}\cos{\theta_{liq}})$ and  $1/(n_{gas}\cos{\theta_{gas}})$, with
$n_{liq} = 1.273$ and $n_{gas} = 1.00194$, are equal. 3 crosses drawn in dark 
blue correspond to anomalous rings coming from the quartz radiator 
(see caption to Fig.~\ref{fig:15}).
The correlation coefficient for the points in this plot is 0.992. The sum of 
$\chi^2$ for the deviations of the 53~points from the diagonal is 40.1. 
The corresponding p-value for the points to be consistent with the diagonal 
is near~95\%.}
\label{fig:16}
\end{figure}

\end{document}